\documentclass[journal=jacsat,manuscript=article]{achemso}

\usepackage[utf8]{inputenc}
\usepackage{physics}
\usepackage{graphicx}
\usepackage{here}
\usepackage{bm}

\usepackage{pdfpages} 
\usepackage{pgffor} 

\makeatletter
\AtBeginDocument{\let\LS@rot\@undefined}
\makeatother

\def\supplementfilename{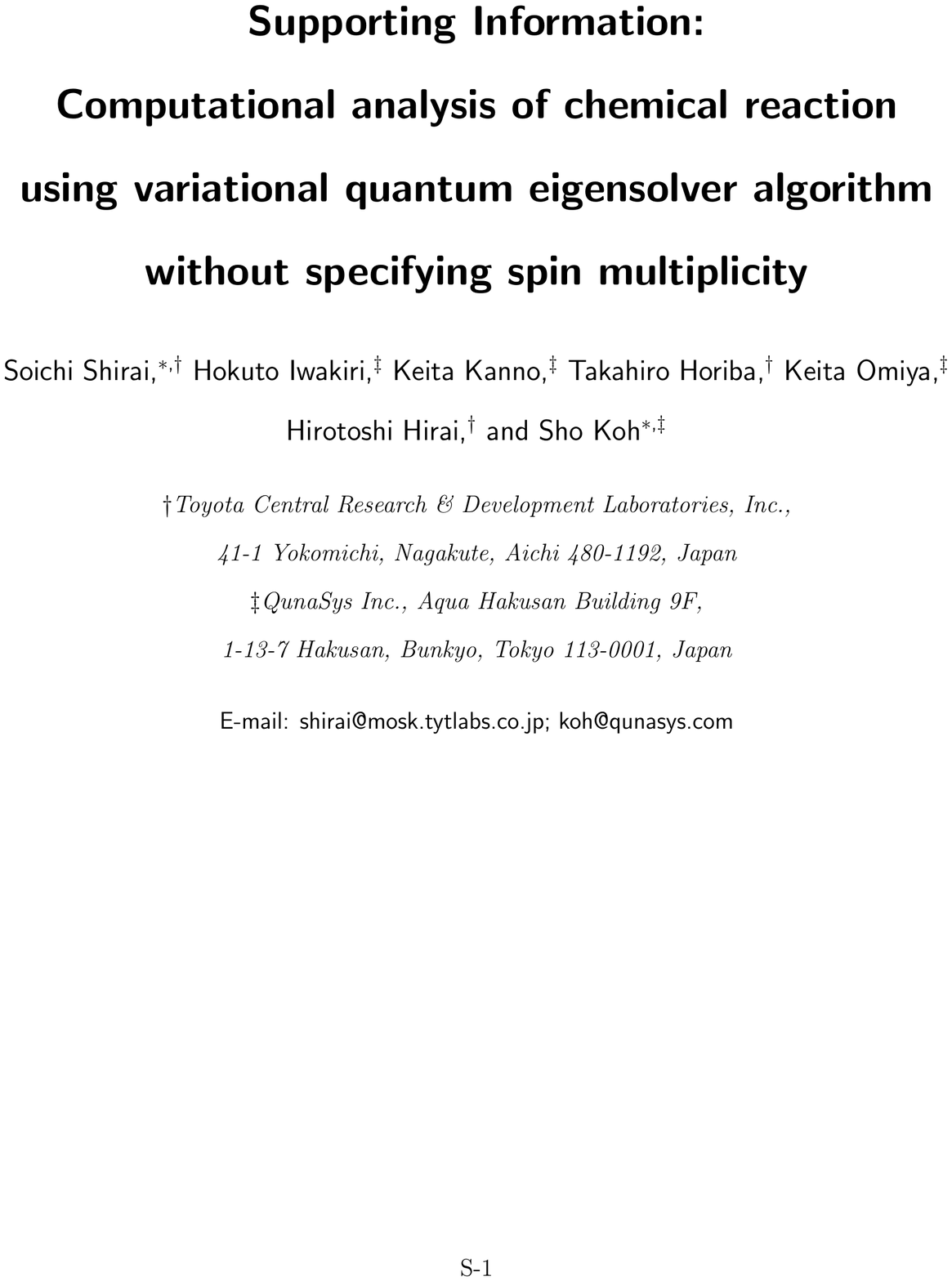}

\pdfximage{\supplementfilename}
\def\numbersupplementpages{\the\pdflastximagepages}

\newif\ifarXiv
\arXivtrue 

\title{Computational analysis of chemical reactions using a variational quantum eigensolver algorithm without specifying spin multiplicity}

\author{Soichi Shirai}
\affiliation[ToyotaCRDL]
{Toyota Central Reserch And Development Laboratories, Incorporated,\\41-1 Yokomichi, Nagakute, Aichi 480-1192, Japan}
\email{shirai@mosk.tytlabs.co.jp}
\author{Hokuto Iwakiri}
\author{Keita Kanno}
\affiliation[QunaSys]
{QunaSys Inc., Aqua Hakusan Building 9F,\\1-13-7 Hakusan, Bunkyo, Tokyo 113-0001, Japan}
\author{Takahiro Horiba}
\affiliation[ToyotaCRDL]
{Toyota Central Reserch And Development Laboratories, Incorporated,\\41-1 Yokomichi, Nagakute, Aichi 480-1192, Japan}
\author{Keita Omiya}
\affiliation[QunaSys]
{QunaSys Inc., Aqua Hakusan Building 9F,\\1-13-7 Hakusan, Bunkyo, Tokyo 113-0001, Japan}

\author{Hirotoshi Hirai}
\affiliation[ToyotaCRDL]
{Toyota Central Reserch And Development Laboratories, Incorporated,\\41-1 Yokomichi, Nagakute, Aichi 480-1192, Japan}
\author{Sho Koh}
\affiliation[QunaSys]
{QunaSys Inc., Aqua Hakusan Building 9F,\\1-13-7 Hakusan, Bunkyo, Tokyo 113-0001, Japan}
\email{koh@qunasys.com}

\begin{document}

\newpage

\section{Abstract}

The analysis of a chemical reaction along the ground state potential energy surface in conjunction with an unknown spin state is challenging because electronic states must be separately computed several times using different spin multiplicities to find the lowest energy state.
However, in principle, the ground state could be obtained with just a single calculation using a quantum computer without specifying the spin multiplicity in advance.
In the present work, ground state potential energy curves for PtCO were calculated as a proof-of-concept using a variational quantum eigensolver (VQE) algorithm. This system exhibits a singlet-triplet crossover as a consequence of the interaction between Pt and CO.
VQE calculations using a statevector simulator were found to converge to a singlet state in the bonding region, while a triplet state was obtained at the dissociation limit.
Calculations performed using an actual quantum device provided potential energies within $\pm2$ kcal/mol of the simulated energies after adopting error mitigation techniques. The spin multiplicities in the bonding and dissociation regions could be clearly distinguished even in the case of a small number of shots. The results of this study suggest that quantum computing can be a powerful tool for the analysis of the chemical reactions of systems for which the spin multiplicity of the ground state and variations in this parameter are not known in advance.

\newpage

\section{Introduction}

Issues related to the use of fossil fuels, including the finite nature of these resources and climate change, have become significant on a global scale.~\cite{fossilfuel2013, climate2021} As such, hydrogen production through water splitting and the conversion of carbon dioxide into liquid fuels have attracted much attention.~\cite{yagi2001, karkas2014, blakemore2015} The development of high-performance catalysts for these processes that contain reduced amounts of noble metals is also required on the basis of economics and sustainability. Quantum chemical calculations are now routinely used to simulate various reactions in conjunction with experimentation to expedite the development of such catalysts by providing an understanding of reaction mechanisms on the molecular level and to assist in material design.

The majority of current industrial catalysts utilize metal or alloy clusters, metal oxides or metal complexes in which the metals serve as active centers~\cite{lloyd2011handbook, hagen2015industrial}. Accordingly, the interactions between adsorbed molecules and these metal atoms is a fundamental aspect of the catalytic reactions and must be accurately described to allow reliable simulations. However, distributing many electrons over the multiple orbitals associated with the catalytic centers and the adsorbed molecules results in numerous electronic states that are close in energy as a consequence of the degenerate or nearly degenerate \textit{d} and \textit{f} orbitals of the metal atoms. Describing such highly correlated electronic states requires a multiconfiguration treatment based on configuration interaction (CI) theory~\cite{ci1999} instead of the more commonly used density functional theory (DFT) methods.

A CI wavefunction is represented as a linear combination of Slater determinants corresponding to the various electron configurations for a system. In the case of so-called full CI, all the electron configurations generated from combinations of all the electrons and orbitals of the system are used~\cite{fci1996}. Although this approach provides exact wavefunctions for an adopted basis set, the number of electron configurations (which, in turn, determines the computational load) rapidly increases along with the number of electrons and orbitals such that the load can easily exceed the capacity of standard computers. Consequently, only truncated CI calculations with selected electron configurations are carried out in practice, except for the simplest molecules. For these reasons, the use of quantum computers for quantum chemical calculations has recently attracted much attention~\cite{qcqc2019, qcqc2020}. Quantum computers with qubits have the potential to perform full CI calculations in polynomial time by employing a quantum-phase estimation (QPE) algorithm to estimate the eigenvalues of specific Hamiltonian operators~\cite{qpe1999, qpe2005, qpe2014, qpe2017}.

In addition to advantages related to reduced computational costs, quantum chemistry calculations performed using quantum computers can also determine electronic ground states without specifying the spin multiplicity in advance.
This is helpful because, in the case of certain transition metal catalysts in which the metals serve as active centers, the spin multiplicity of the electronic ground state varies depending on the molecular geometry~\cite{kitagawa2016dft, watanabe2016spin, jiang2016mechanism, kwon2017catalytic, Nakatani2018, li2020identification}.
Thus, it is necessary to trace changes in the spin multiplicity along the reaction coordinates when assessing such systems.
In contrast, the spin multiplicity must be specified when performing DFT calculations.
For CI calculations using conventional computers, in order to save the computational costs, the spin multiplicity is also specified in advance to omit the electron configurations which do not contribute to the target spin state\cite{Christopher2004}.
For this reason, the analysis of a chemical reaction along the ground state potential energy surface is challenging because it is necessary to compute electronic states several times while specifying different spin multiplicities and to compare the resulting energies to identify the lowest energy state.
Conversely, since a superposition of multiple states having different spin multiplicities can be prepared using the qubits of a quantum computer, the ground state can, in principle, be obtained with just a single calculation employing the QPE algorithm if the initial state overlaps significantly with the ground state.
As such, it will be possible to trace the reaction path along the ground state even if the spin multiplicity changes along the way without involving complicated operations, unlike aforementioned calculations performed with conventional computers.
Therefore, quantum computers are expected to serve as a useful tool for the analysis of the reactions of a system for which the spin multiplicity of the ground state is not obvious. This computational technique could also be applied to analyses of the ground states of strongly correlated systems having electronic states that are close in energy to the ground state but different in terms of spin multiplicity~\cite{radon2008binding, sebetci2009does, ali2012electronic, garcia2017effect, zhang2020probing}.
Unfortunately, such QPE calculations require fault tolerance and cannot be executed on current quantum computers having qubits without error correction, generally referred to as noisy intermediate-scale quantum (NISQ) devices~\cite{nisq2018, nisq2019, nisq2021}. 
Accordingly, a quantum-classical hybrid algorithm known as the variational quantum eigensolver (VQE) is widely employed for quantum chemical calculations using NISQ devices~\cite{vqe2014, vqe2016}.
These VQE calculations solve the ground state wavefunction by minimizing a given cost function based on the variational principle.
In the case of a VQE scheme, adopting an ansatz describing multiple spin states can allow the ground state to be directly determined by minimizing the cost function.

The present study demonstrates that a reaction path along a ground state potential energy curve including spin crossover can be traced using quantum algorithms. Specifically, the potential energy curves for the adsorption of a carbon monoxide (CO) molecule on a platinum (Pt) atom were calculated using a VQE algorithm as a proof-of-concept. This adsorption is one of the most extensively studied systems in the fields of surface and catalytic chemistry and PtCO is the simplest model for this reaction. Prior theoretical studies have suggested that this system has a singlet ground state in its equilibrium geometry~\cite{ptco1993, ptco2004} but a triplet state at the dissociation limit because of the triplet ground state of the neutral Pt atom~\cite{platinum1992}.
Thus, the ground state of PtCO exhibits spin crossover resulting from the interaction between the Pt and CO. 
In this work, VQE calculations were used to calculate the ground state potential energy curves of PtCO, including a singlet-triplet spin crossover, without specifying the spin multiplicity in advance.
The spin multiplicity could be estimated efficiently based on a small number of measurements using an actual quantum device even under noisy conditions, implying that the evaluation of discrete values may be a useful application of NISQ devices.

\section{Theory}
\subsection{VQE}
The VQE method is a quantum-classical hybrid technique based on the variational principle~\cite{vqe2014, vqe2016}.
In this process, a trial wavefunction is constructed using a quantum computer based on the initial wavefunction $\ket{\Psi_o}$ and the ansatz quantum circuit $\hat{U}(\bm{\theta})$ with variational parameters $\theta$. 
The expected value of the molecular Hamiltonian $\bra{\Psi_o} \hat{U}^\dagger(\bm{\theta}) \hat{H} \hat{U}(\bm{\theta})\ket{\Psi_o}$ is repeatedly determined with a quantum computer, while the variational parameters $\bm{\theta}$ are repeatedly updated using a classical computer so as to minimize the cost function 
\begin{equation}
    \textit{L$_{cost}$}=\bra{\Psi_o} \hat{U}^\dagger(\bm{\theta}) \hat{H}\hat{U}(\bm{\theta})\ket{\Psi_o}.
\end{equation}
To ensure that the computed wavefunction has the desired properties, a penalty term is typically included as~\cite{vqe2016, rvqe2019, kuroiwa2021} 
\begin{equation}
    \textit{L$_{cost}$}=\bra{\Psi_o} \hat{U}^\dagger(\bm{\theta}) \hat{H}\hat{U}(\bm{\theta})\ket{\Psi_o}+\textit{L$_{penalty}$}.
\end{equation}
In the case of spin multiplicity, \textit{L$_{penalty}$} is typically written as
\begin{equation}
    \textit{L$_{penalty}$}=\textit{w}\bra{\Psi_o} \hat{U}^\dagger(\bm{\theta}) (\hat{S}^{2}-\textit{p})^{2}\hat{U}(\bm{\theta})\ket{\Psi_o},
\end{equation}
where \textit{w} is a weighting coefficient that determines the magnitude of the penalty, $\hat{S^2}$ is the square of the total spin angular momentum
operator, and \textit{p} is the eigenvalue of $\hat{S^2}$ that the wavefunction should satisfy.  Note that the spin multiplicity can be specified by defining \textit{p}. In contrast, the wavefunction can be calculated without specifying the spin multiplicity by setting \textit{w} to zero. This computational scheme was utilized in the present study.

\subsection{PtCO}
\begin{figure}[H]
    \begin{center}
        \includegraphics[width=15.0cm]{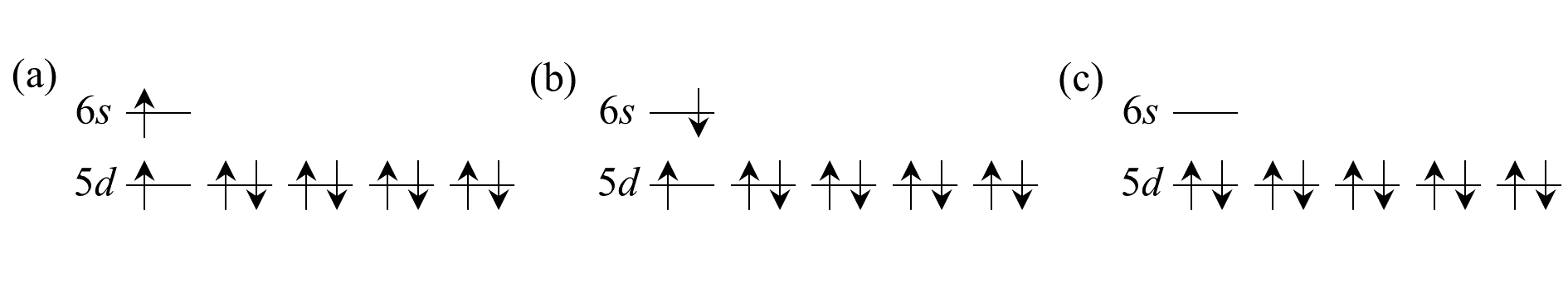}
        \caption{Diagrams showing the electron configurations associated with the (a) $^{3}$D, (b) $^{1}$D, and (c) $^{1}$S states of a neutral Pt atom.}
        \label{3D1D1S}
    \end{center}
\end{figure}

\begin{figure}[H]
    \begin{center}
        \includegraphics[width=12.0cm]{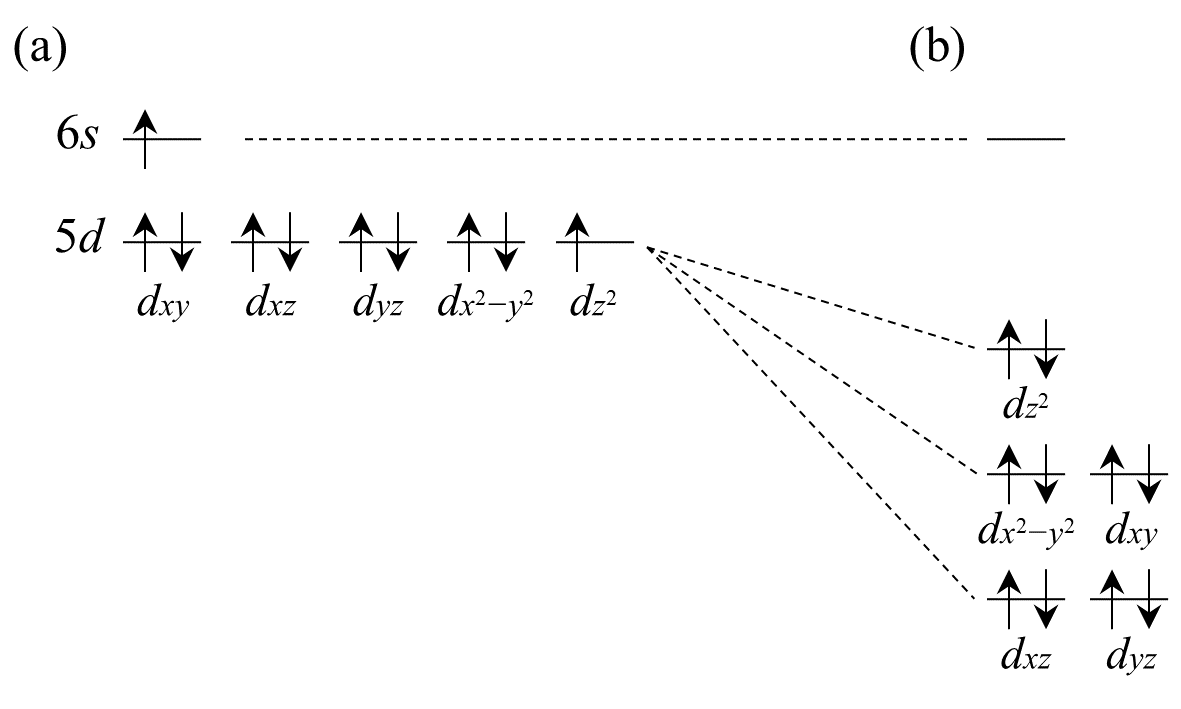}
        \caption{Diagrams showing the alignments of the molecular orbitals of PtCO derived from Pt 5\textit{d} and 6\textit{s} atomic orbitals for the (a) dissociation limit and (b) bonding region. A PtCO molecule that is linear along its \textit{z}-axis is assumed.}
        \label{MO-split}
  \end{center}
\end{figure}

This work focused on the low-lying singlet and triplet states of PtCO, either of which could be the ground state. These states could be identified based on the electron configurations of the molecular orbitals derived from the Pt 5\textit{d} and 6\textit{s} atomic orbitals. The ground state of a neutral Pt atom is the triplet state, equivalent to the electron configuration (5\textit{d})$^{9}$(6\textit{s})$^{1}$, and this state is denoted herein by the symbol $^{3}$D. The lowest energy singlet state, $^{1}$D, has the same electron configuration as $^{3}$D but a total spin angular momentum of $S = 0$. In contrast, the $^{1}$S state is associated with the closed-shell electron configuration (5\textit{d})$^{10}$(6\textit{s})$^{0}$. The electron configurations of these states are illustrated in Figure \ref{3D1D1S}. The energy levels of these states increase in the order of $^{3}$D \textless\ $^{1}$D \textless\ $^{1}$S. Therefore, the ground state of PtCO at the dissociation limit is a triplet state derived from the $^{3}$D state of Pt. The $^{3}$D and $^{1}$D states are both five-fold degenerate since all five \textit{d} orbitals have the same energy. However, in the case of the PtCO model, the degeneracy of the \textit{5d} orbitals is removed as a result of interactions with the CO molecule. Assuming that PtCO has a linear structure, as previously suggested by a DFT study performed by Wu et al.,~\cite{ptco2004} the Pt 5\textit{d} orbitals will be split into three energy levels as shown in Figure \ref{MO-split}. The energy gap between the Pt 6\textit{s} orbital and the highest Pt 5\textit{d} orbital is therefore increased by the presence of the CO. In the case that the energy gap between the Pt 5\textit{d} and 6\textit{s} orbitals is greater than the value for the exchange interaction, a closed-shell singlet state having a (5\textit{d})$^{10}$(6\textit{s})$^{0}$ configuration becomes energetically preferable to the triplet state with a (5\textit{d})$^{9}$(6\textit{s})$^{1}$ configuration. Therefore, the ground state of PtCO can comprise a closed-shell singlet state in the case that the Pt and CO are in close proximity to one another.

\newpage

\newpage

\newpage

\section{Computational details}
Conventional quantum chemical calculations were initially performed to determine the computational procedure and conditions required for the VQE calculations. Subsequent to this, VQE calculations were performed using both a simulator and a quantum device to demonstrate that the ground state could be determined without specifying the spin multiplicity. In preparation for these calculations, the geometry of the PtCO molecular structural was optimized using the DFT method with the B3LYP functional~\cite{b3lyp1980, b3lyp1988, b3lyp1993, b3lyp1994} together with the def2-QZVP basis set for the Pt atom~\cite{ahlrichs2005} and the cc-pVQZ basis set for the C and O atoms~\cite{dunning1989}. The Gaussian16 program was employed for these calculations~\cite{g16} and a linear-structured PtCO with an \textit{r}(C--O) value of 1.1446 Å was obtained. Note that, herein, \textit{r}(X--A) represents the interatomic distance between atoms X and A. The \textit{r}(C--O) distance was fixed at this value in the following calculations.
Plots of potential energy as a function of \textit{r}(Pt--C) were generated over the \textit{r}(Pt--C) range of 1.55 -- 5.0 Å. In these calculations, the linear PtCO molecule was placed on the \textit{z}-axis and \textit{C}$_{2\textit{v}}$ symmetry was applied. Accordingly, the symmetries of the \textit{d$_{xy}$}, \textit{d$_{yz}$}, \textit{d$_{xz}$}, \textit{d$_{x^{2}-y^{2}}$} and \textit{d$_{z^{2}}$} orbitals were \textit{a}$_{2}$, \textit{b}$_{2}$, \textit{b}$_{1}$, \textit{a}$_{1}$, and \textit{a}$_{1}$, respectively, while that of the molecular orbital derived from Pt 6\textit{s} was \textit{a}$_{1}$. Since a closed-shell singlet state has \textit{A}$_{1}$ symmetry, those electronic states having \textit{A}$_{1}$ symmetry were investigated in this study. It should also be noted that spin-orbit (SO) interactions were not incorporated into the calculations. Although SO interactions typically have a significant effect in systems that include heavy elements such as Pt, past theoretical studies have indicated that the singlet-triplet crossover in the ground state potential energy curve of PtCO can be reproduced without considering SO effects~\cite{ptco1993}. In the present work, a qualitative description of the spin crossover was sufficient to allow the applicability of the quantum algorithm to these calculations without specifying the spin multiplicity to be ascertained.

\subsection{Calculations with conventional methods}
The potential energy curves for PtCO were calculated using the complete active space self-consistent field (CASSCF) method~\cite{casscf1980, casscf1987} and the complete active space configuration interaction (CASCI) method~\cite{casci2021}. The former approach is widely used to determine electronic states associated with multiple electron configurations. In CASSCF calculations, the molecular orbitals and the expansion coefficients of electron configurations are both optimized for the target state. Calculations at the same level of theory can be performed within the VQE framework by adopting an orbital optimization scheme (OO-VQE)~\cite{oovqe2020, oovqe2022} although this requires a huge number of calculations because of the iterations involved in the VQE process, leading to a high computational cost. Therefore, the present work did not employ the orbital optimization technique. The accuracy of the CASCI calculations was also examined based on a comparison with the CASSCF results. The present CASSCF and CASCI calculations were carried out using the GAMESS program.~\cite{gamess1993, gamess2005}

\subsubsection{CASSCF calculations}
The accuracy of CASSCF calculations depends on the construction of the active space, and this work employed the CAS(10e, 6o), CAS(4e, 3o) and CAS(2e, 2o) active spaces. Herein, the CAS is denoted as CAS(\textit{M}e, \textit{N}o) based on the numbers of active electrons \textit{M} and of active orbitals \textit{N}. The CAS(10e, 6o) space was constructed by distributing ten electrons over six molecular orbitals derived from Pt 5\textit{d} and 6\textit{s} atomic orbitals. The CAS(4e, 3o) space was constructed by distributing four electrons over three orbitals having \textit{a}$_{1}$ symmetry selected from the active orbitals of the CAS(10e, 6o) space. Finally, the CAS(2e, 2o) space was obtained by selecting the two highest energy orbitals from the active orbitals of the CAS(4e, 3o) space. The three lowest energy singlet states and the two lowest energy triplet states having \textit{A}$_{1}$ symmetry were evaluated in the calculations based on the CAS(10e, 6o) and CAS(4e, 3o) active spaces. Henceforth, these states are denoted as 1$^{1}$\textit{A}$_{1}$, 2$^{1}$\textit{A}$_{1}$, 3$^{1}$\textit{A}$_{1}$, 1$^{3}$\textit{A}$_{1}$ and 2$^{3}$\textit{A}$_{1}$. When using the CAS(2e, 2o) space, the two lowest energy singlet states and the lowest triplet state were calculated. The basis sets were def2-QZVP for the Pt atom and cc-pVQZ for the C and O atoms.

\subsubsection{CASCI calculations}
The potential energy curves for PtCO associated with the two lowest energy singlet states and the lowest triplet state were calculated using the CASCI method with the CAS(2e, 2o) space. It should be noted that the results were dependent on the molecular orbitals that were considered because orbital optimization was neglected. Restricted Hartree-Fock (RHF) orbitals and restricted open-shell HF (ROHF) orbitals were both examined. The ROHF orbitals were obtained for the lowest energy triplet state, 1$^{3}$\textit{A}$_{1}$. The effects of the basis set were also evaluated, employing the def2-SVP, def2-TZVP, and def2-QZVP basis sets for the Pt atom combined with the cc-VDZ, cc-pVTZ, and cc-pVQZ basis sets for the C and O atoms, respectively.

\subsection{VQE calculations using a simulator and a quantum device}
\begin{figure}[H]
    \begin{center}
        \includegraphics[width=9.0cm]{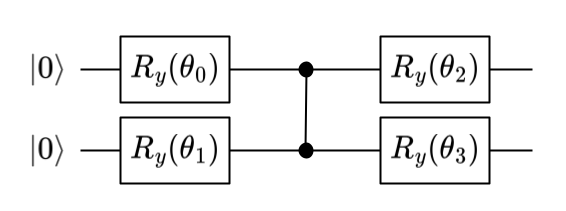}
        \caption{A diagram of the present ansatz consisting of $R_y$ rotation gates and a $CZ$ entanglement gate. The computational basis $\ket{00}$ was employed as the initial state and four variational parameters $\theta_i$ ($i=0,1,2,3$) for the $R_y$ rotation gates were optimized during the VQE calculations.}
        \label{Ry_ansatz}
    \end{center}
\end{figure}

The potential energy curves and the total spin angular momentum values for PtCO were calculated using the VQE method without penalty terms, employing both the Qiskit simulator~\cite{Qiskit} and the IBM \textit{ibm\_canberra} quantum device ~\cite{ibm_quantum}.
The basis sets used were def2-SVP for the Pt atom and cc-pVDZ for the C and O atoms.
The molecular orbitals were prepared using the ROHF method implemented in PySCF~\cite{sun2018pyscf}.
The same active orbitals together with the CAS(2e, 2o) state as used in the CASCI calculations were adopted.
Parity transformation was used to map the wavefunction to qubits and reduce two qubits based on parity conservation of the electron number of $\alpha$ spin and the total electron number. As shown in Figure \ref{Ry_ansatz}, a hardware-efficient ansatz~\cite{he2017} omitting $R_z$ gates with a single depth was applied to the computational basis $\ket{00}$ to describe the target wavefunction in real space. The variational parameters $\theta_i$ ($i=0,1,2,3$) for $R_y$ gates, starting from random initial numbers, were optimized using the Nakanishi-Fujii-Todo (NFT) optimizer~\cite{NFT}.

Simulator calculations were carried out to ascertain whether the VQE algorithm itself could provide the ground state without specifying the spin multiplicity. The noise effect was ignored and the potential energy and total spin angular momentum with statevector were calculated using conventional computers. 
In the case of those calculations using the \textit{ibm\_canberra} machine, the potential energy and total spin angular momentum of PtCO at \textit{r}(Pt--C) values of 1.846 and 3.0 Å were determined through the Qiskit Runtime~\cite{ibm_quantum}. 
The \textit{r}(Pt--C) distance of 1.846 Å was chosen because the ground state potential energy curve was found to exhibit its minimum at this point, whereas 3.0 Å was selected as a typical bond length for a triplet ground state.
Three VQE calculations were performed for each molecular structure with a maximum of five iterations for the VQE optimization so as to reduce the computational cost.
The Hamiltonian operator was decomposed into seven Pauli operators that were then placed into four groups using a bitwise grouping technique~\cite{vqe2016} to reduce the number of measurements. These measurements were performed using 5008 shots for each group of Pauli operators to ascertain energies during the VQE process. The square of the total spin angular momentum operator $\hat{S^2}$, which consists of the three Pauli operators to be determined, was ascertained for the optimized state, employing 16 shots for each group.
The twirled readout error extinction (T-Rex) method~\cite{van_den_Berg_2022} implemented in the IBM Qiskit Runtime was adopted to reduce the readout noise effect. 

\newpage

\newpage

\section{Results and Discussions}
\subsection{CASSCF and CASCI calculations}
\begin{figure}[H]
    \begin{center}
        \includegraphics[scale=1.0]{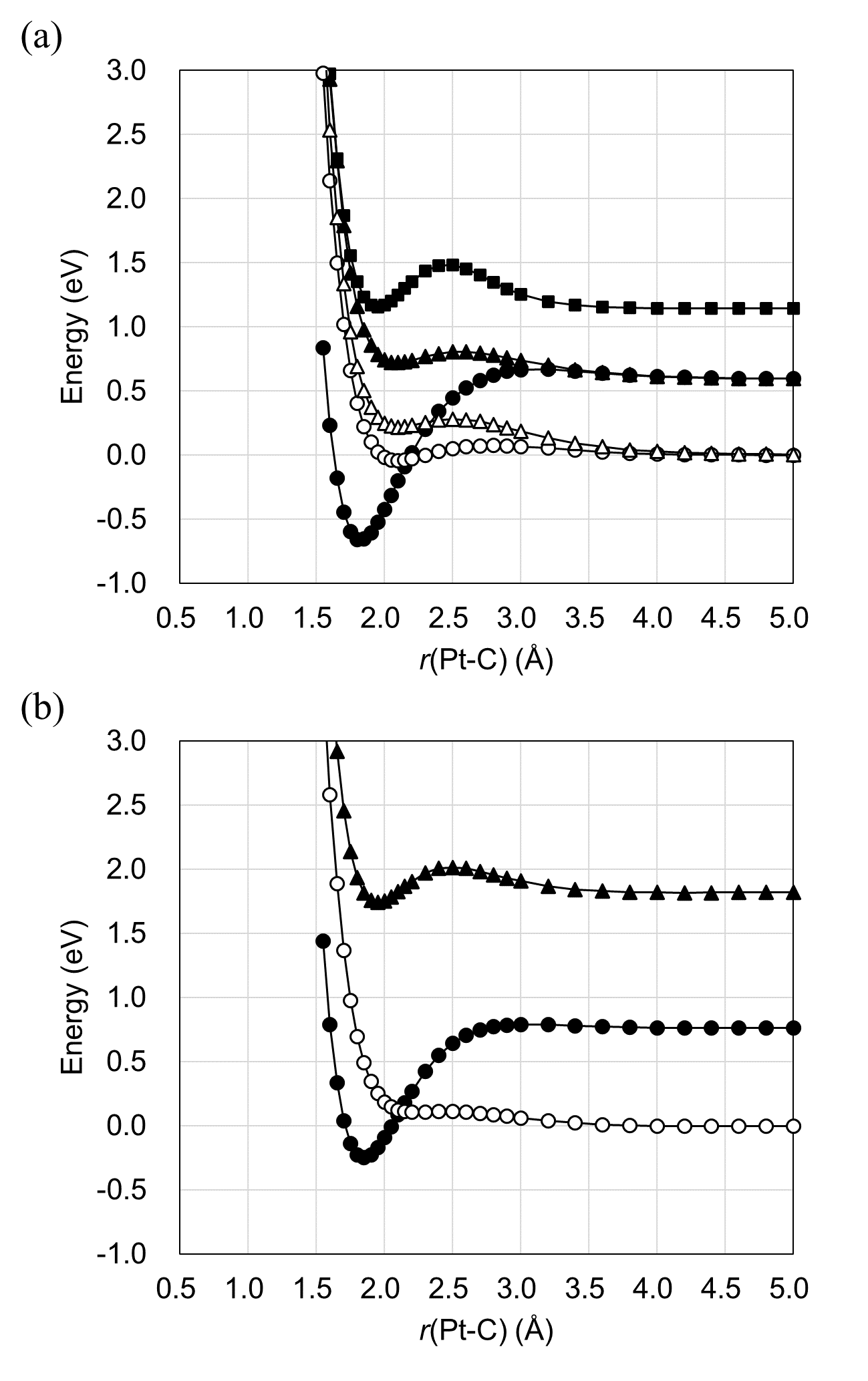}
        \caption{Potential energy curves for the 1$^{1}$\textit{A}$_{1}$ (filled circles), 2$^{1}$\textit{A}$_{1}$ (filled triangles), 3$^{1}$\textit{A}$_{1}$ (filled squares), 1$^{3}$\textit{A}$_{1}$ (open circles), and 2$^{3}$\textit{A}$_{1}$ (open triangles) states as calculated using (a) the CASSCF method with the CAS(10e, 6o) space and (b) the CASCI method based on ROHF orbitals. The basis sets used in the CASSCF calculations were def2-QZVP for Pt and cc-pVQZ for C and O, while those in the CASCI calculations were def2-SVP for Pt and cc-pVDZ for C and O.}
        \label{casscf-casci}
    \end{center}
\end{figure}

The total energies calculated for the electronic states are summarized in Tables S1--S5.
The potential energy curves obtained using the CASSCF method with the CAS(10e, 6o) state are presented in Figure \ref{casscf-casci}(a). Herein, it is assumed that, at \textit{r}(Pt--C) = 5.0 Å, the structure is at its dissociation limit. At this limit, the doubly degenerate triplet states 1$^{3}$\textit{A}$_{1}$ and 2$^{3}$\textit{A}$_{1}$ are the lowest energy states, while the doubly degenerate singlet states 1$^{1}$\textit{A}$_{1}$ and 2$^{1}$\textit{A}$_{1}$ are the second lowest, and the third singlet state 3$^{1}$\textit{A}$_{1}$ is the highest. An analysis of the CASSCF wavefunction suggests that these triplet states are derived from the $^{3}$D state of Pt having the electron configuration (5\textit{d})$^{9}$(6\textit{s})$^{1}$ (Figure S1). Similarly, the 1$^{1}$\textit{A}$_{1}$ and 2$^{1}$\textit{A}$_{1}$ states are derived from the $^{1}$D state with the configuration (5\textit{d})$^{9}$(6\textit{s})$^{1}$, while the 3$^{1}$\textit{A}$_{1}$ state originates from the $^{1}$S state with the configuration (5\textit{d})$^{10}$(6\textit{s})$^{0}$. The energetic ordering of these states is consistent with the energy levels of neutral Pt~\cite{platinum1992}. The normally degenerate states are split with regard to their energy levels due to the interactions of the Pt atom with the CO. Consequently, the energetic ordering of the 1$^{1}$\textit{A}$_{1}$ and 1$^{3}$\textit{A}$_{1}$ states is inverted at \textit{r}(Pt--C) = 2.17 Å and the ground state in the region over which \textit{r}(Pt--C) $<$ 2.17 Å is 1$^{1}$\textit{A}$_{1}$. Thus, the spin multiplicity of the ground state varies depending on the value of \textit{r}(Pt--C). The potential energy curves obtained using the CASCI method based on ROHF orbitals are provided in Figure \ref{casscf-casci}(b). Although the energy level of the 1$^{3}$\textit{A}$_{1}$ state is relatively underestimated here compared with the CASSCF results, the crossover of the 1$^{1}$\textit{A}$_{1}$ state with 1$^{3}$\textit{A}$_{1}$ was also predicted at this level of theory and appears at \textit{r}(Pt--C) = 2.12 Å. These results imply that VQE calculations using fewer qubits could potentially describe the spin crossover in the ground state.

\begin{figure}[H]
    \begin{center}
        \includegraphics[scale=1.5]{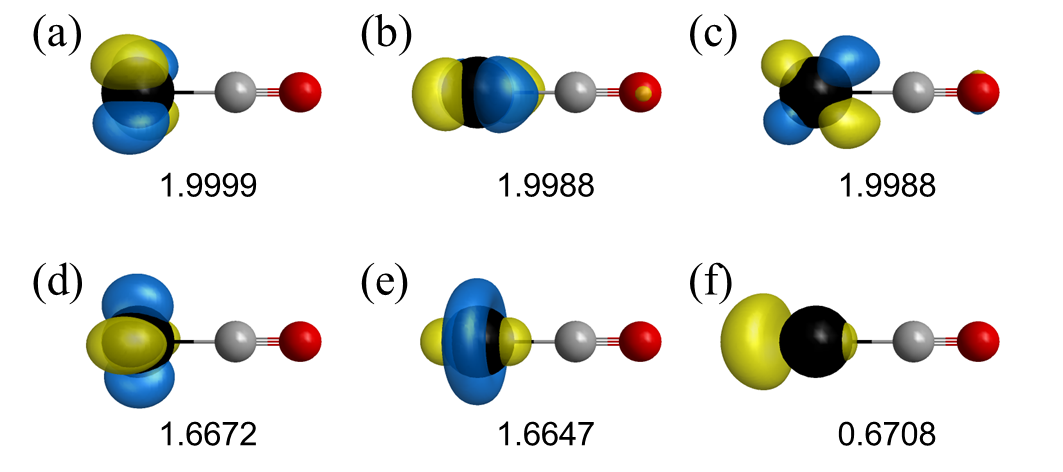}
        \caption{Natural orbitals at \textit{r}(Pt--C) = 1.85 Å for the singlet states calculated using the CASSCF method with the CAS(10e, 6o) space. These orbitals are derived from the (a) 5\textit{d$_{xy}$}, (b) 5\textit{d$_{yz}$}, (c) 5\textit{d$_{xz}$}, (d) 5\textit{d$_{x^{2}-y^{2}}$}, (e) 5\textit{d$_{z^{2}}$}, and (f) 6\textit{s} atomic orbitals of Pt. The associated symmetries are \textit{a}$_{2}$, \textit{b}$_{2}$, \textit{b}$_{1}$, \textit{a}$_{1}$, \textit{a}$_{1}$, and \textit{a}$_{1}$, respectively. The occupation numbers are shown below the orbitals.}
        \label{natorb185}
    \end{center}
\end{figure}

An analysis of the CASSCF wavefunctions with the CAS(10e, 6o) space suggested that the main configuration of the 1$^{3}$\textit{A}$_{1}$ state was (5\textit{d})$^{9}$(6\textit{s})$^{1}$ with a singly occupied 5\textit{d$_{z^{2}}$} orbital over the entire region, while that for the 2$^{1}$\textit{A}$_{1}$ and 2$^{3}$\textit{A}$_{1}$ states was (5\textit{d})$^{9}$(6\textit{s})$^{1}$ with a singly occupied \textit{d$_{x^{2}-y^{2}}$} orbital. In contrast, the weights of the main configurations for the 1$^{1}$\textit{A}$_{1}$ and 3$^{1}$\textit{A}$_{1}$ states varied depending on \textit{r}(Pt--C) (Figure S1). Here, the weights were calculated as squared coefficients of the configurations in the CASSCF wavefunctions. The main configurations for these states were (5\textit{d})$^{9}$(6\textit{s})$^{1}$ with a singly occupied 5\textit{d$_{z^{2}}$} orbital together with the closed-shell (5\textit{d})$^{10}$(6\textit{s})$^{0}$ configuration. For the 1$^{1}$\textit{A}$_{1}$ state, the configuration (5\textit{d})$^{9}$(6\textit{s})$^{1}$ was dominant at the dissociation limit, while (5\textit{d})$^{10}$(6\textit{s})$^{0}$ was dominant in the bonding region. Conversely, in the case of the 3$^{1}$\textit{A}$_{1}$ state, the configuration (5\textit{d})$^{10}$(6\textit{s})$^{0}$ was dominant at the dissociation limit, but (5\textit{d})$^{9}$(6\textit{s})$^{1}$ was the primary configuration throughout the bonding region. The total weights of these configurations were greater than 0.85 for all states over the entire region. Figure \ref{natorb185} shows the CASSCF natural orbitals  and the associated occupation numbers at \textit{r}(Pt--C) = 1.85 Å on behalf of the structures  in the bonding region. The sum of the occupation numbers for the orbitals with \textit{a}$_{1}$ symmetry was nearly four, while the other orbitals were approximately doubly occupied. Similar results were obtained for the dissociation limit, as shown in Figure S2. Therefore, the singlet--triplet crossover was predicted reasonably well using the smaller active spaces CAS(4e, 3o) and CAS(2e, 2o) (Figure S3).

The results of CASCI calculations based on the RHF orbitals were found to be inconsistent with those obtained using the CASSCF method (Figure S4). Specifically, the energetic ordering of the electronic states was 1$^{1}$\textit{A}$_{1}$ $<$ 1$^{3}$\textit{A}$_{1}$ over the entire region and a 1$^{1}$\textit{A}$_{1}$--1$^{3}$\textit{A}$_{1}$ crossover was not observed. Because the RHF orbitals were optimized for the closed-shell singlet state, the energy level of the electronic state having the primary configuration (5\textit{d})$^{10}$(6\textit{s})$^{0}$ could have been significantly underestimated. An analysis of the CASCI wavefunction revealed that one of the main configurations of 1$^{1}$\textit{A}$_{1}$ was derived from the (5\textit{d})$^{10}$(6\textit{s})$^{0}$ configuration of the Pt atom, while that of the 1$^{3}$\textit{A}$_{1}$ state originated from a (5\textit{d})$^{9}$(6\textit{s})$^{1}$ configuration. Thus, the energy level of the 1$^{1}$\textit{A}$_{1}$ state was relatively underestimated, resulting in the absence of the 1$^{1}$\textit{A}$_{1}$--1$^{3}$\textit{A}$_{1}$ crossover. In contrast, in the case of the CASCI calculations based on ROHF orbitals, the energy level of the 1$^{3}$\textit{A}$_{1}$ state was underestimated relative to those of the 1$^{1}$\textit{A}$_{1}$ state because the orbitals were optimized for the lowest triplet state. Consequently, the 1$^{1}$\textit{A}$_{1}$--1$^{3}$\textit{A}$_{1}$ crossover was observed. The effect of the basis set was found to be minimal and the 1$^{1}$\textit{A}$_{1}$--1$^{3}$\textit{A}$_{1}$ crossover could be reproduced even with smaller basis sets (Figure S5). Therefore, the combination of def2-SVP for Pt and cc-pVDZ for C and O, which were the smallest basis sets among those examined, was adopted for the VQE calculations to reduce the computational load.
\newpage
\newpage
\newpage

\subsection{VQE calculations}
\begin{figure}[H]
    \begin{center}
        \includegraphics[width=10.5cm]{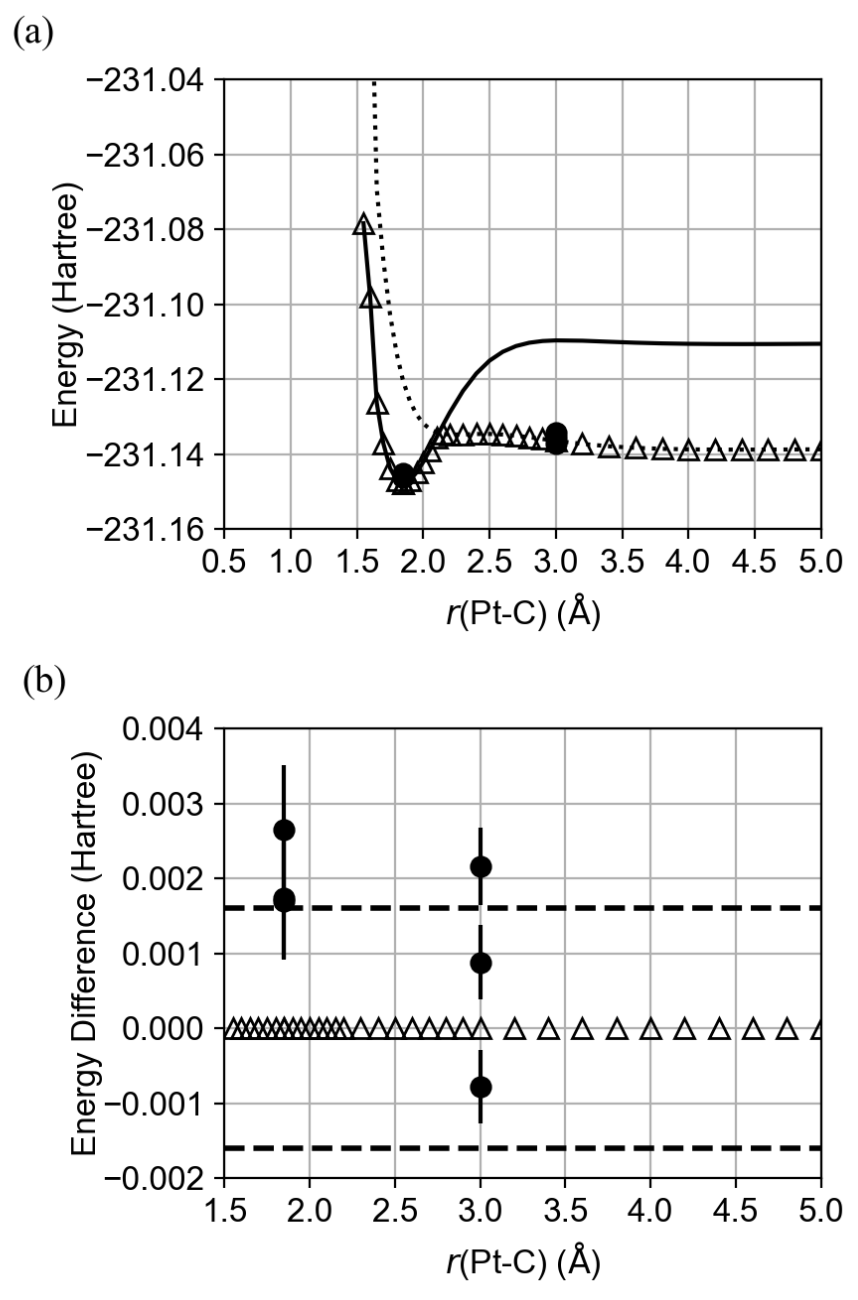}
        \caption{(a) Potential energy curves of the singlet ground state (solid line) and the triplet ground state (dotted line) calculated using the CASCI method, the potential energies obtained using the VQE approach with a statevector simulator (open triangles) and the IBM \textit{ibm\_canberra} quantum device (filled circles). (b) The energy difference from the CASCI potential energies obtained from the simulator (open triangles) and the IBM \textit{ibm\_canberra} quantum device (filled circles). The range indicated by the two dashed lines equals $\pm 0.0016$ Hartree, equivalent to so-called chemical accuracy ($\pm 1$ kcal/mol).}
        \label{pec_sim_exp}
    \end{center}
\end{figure}

The potential energy curves calculated using the VQE technique with the statevector simulator are plotted in Figure \ref{pec_sim_exp}(a) along with the CASCI results.
These VQE calculations with the simulator reproduced the singlet potential energy curve obtained from the CASCI process over the range of \textit{r}(Pt--C) $<$ 2.12 Å and the triplet potential energy curve over the range of \textit{r}(Pt--C) $>$ 2.12 Å. Thus, the VQE values follow the ground state potential energy curve over the entire region.
These results suggest that the VQE algorithm is able to closely predict a reaction pathway along a ground state potential energy curve including a spin crossover.
The potential energies at \textit{r}(Pt--C) = 1.846 Å and 3.0 Å were also calculated using the IBM quantum device and the resulting energies and the associated errors obtained from three replicate VQE calculations are collected in Table \ref{tab:ibm_result}.
These results qualitatively reproduce the CASCI data as plotted in Figure \ref{pec_sim_exp}(a). The calculated values were close to the singlet potential energy curve generated via the CASCI method for \textit{r}(Pt--C) = 1.846 Å and the triplet potential energy curve obtained from the CASCI technique for \textit{r}(Pt--C) = 3.0 Å.
The differences between the potential energy values acquired from the IBM quantum device and the exact values calculated using the CASCI method are plotted in Figure \ref{pec_sim_exp}(b).
It is evident from this plot that these differences were all less than $\pm{0.003}$ Hartree ($\pm{2}$ kcal/mol).

\begin{figure}[H]
    \begin{center}
        \includegraphics[width=10.5cm]{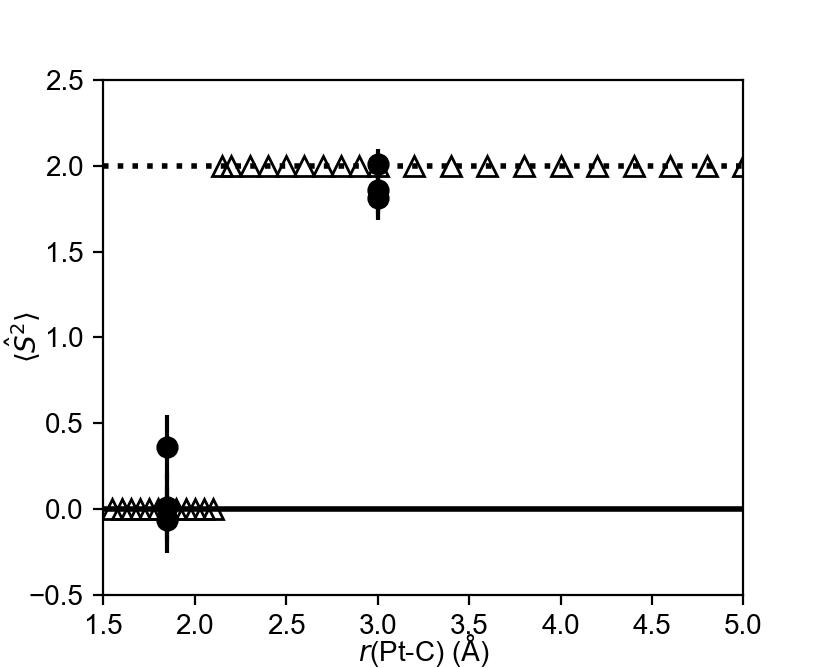}
        \caption{Expected values of the spin squared operator $\ev*{\hat{S^2}}$ as obtained from the simulator (open triangles) and the IBM \textit{ibm\_canberra} quantum device (filled circles). The solid and dotted lines show the expected values of the spin squared operator $\ev*{\hat{S^2}}$, which are 0.0 and 2.0 for the singlet and triplet states, respectively.}
        \label{ibm_spin}
    \end{center}
\end{figure}

The calculated values for the spin squared operator, $\ev*{\hat{S^2}}$, are also summarized in Table \ref{tab:ibm_result} and plotted in Figure \ref{ibm_spin}.
All the calculated $\ev*{\hat{S}^{2}}$ values were within $\pm0.5$ of the reference values calculated using the CASCI technique.
Consequently, the singlet and triplet states could be clearly distinguished when these values were rounded off to the nearest integers.
The errors in these results are sufficiently small such that the change in spin multiplicity can be recognized even though only $16$ shots were carried out for each Pauli term in $\hat{S}^{2}$.
It is known that the minimal difference in the $\ev*{\hat{S}^{2}}$ values between the electronic states having different spin multiplicities will be 2.0 for a system with an even number of electrons, e.g. the $\ev*{\hat{S}^{2}}$ values are 0.0 and 2.0 for singlet and triplet states, respectively.
This value will be 3.0 for a system with an odd number of electrons, e.g. the $\ev*{\hat{S}^{2}}$ values are 0.75 and 3.75 for doublet and quartet states, respectively. Accordingly, the change in spin multiplicity can be detected even in the case of results that have a moderate level of accuracy when the parity of the electron number is conserved.
Since the standard error associated with the expected value of the Pauli term $\hat{P}$ is $\sqrt{1-\ev*{\hat{P}}^2} / \sqrt{N}$, where $N$ is the shot number, approximately 10 shots are sufficient to obtain the required precision in the ones place for a single Pauli term in the absence of noise.
The spin squared operator used in these calculations, $\hat{S^2}$, is expressed as 
\begin{equation}
    \hat{S}^{2} = -0.5 \hat{X_0} \hat{X_1} + 0.5 \hat{Y_0} \hat{Y_1} + 0.5 \hat{Z_0} \hat{Z_1} + 0.5,
\end{equation}
where $\hat{X_i}, \hat{Y_i}$ and $\hat{Z_i}$ denote the Pauli operators for qubit $i$ $(i=0, 1)$ along the $x, y, z$ directions, respectively.
Here we can let $e_P$ equal the largest standard error among the standard errors of $\ev*{\hat{X_0} \hat{X_1}}$, $\ev*{\hat{Y_0} \hat{Y_1}}$ and $\ev*{\hat{Z_0} \hat{Z_1}}$ using $N$ shots.
Based on the propagation of error, the standard error of $\ev*{\hat{S}^{2}}$ is then
\begin{eqnarray}
    e_{S^2} &=& \sqrt{(-0.5)^2 (1 - \ev*{\hat{X_0} \hat{X_1}}^2) / N + 0.5^2 (1 - \ev*{\hat{Y_0} \hat{Y_1}}^2) / N + 0.5^2 (1 - \ev*{\hat{Z_0} \hat{Z_1}}^2) / N} \nonumber \\
                 &\leq& \sqrt{(-0.5)^2 e_P^2 + 0.5^2 e_P^2 + 0.5^2 e_P^2} = \sqrt{0.75} e_P  \nonumber \\
                 &<& e_P.
\end{eqnarray}
Therefore, the use of ten shots for a single Pauli term is sufficient to obtain the necessary numeric precision in the ones place for $\ev*{\hat{S}^{2}}$ in these calculations.
On this basis, the spin multiplicity of the ground state could be successfully calculated via the VQE process even with present-day noisy devices because the determination of the square of the total spin angular momentum is essentially unaffected by noise.

The total spin angular momentum $\hat{S}$ can be determined with a smaller number of measurements than are required for $\hat{H}$. This can be explained based on the number of Pauli terms that appear in the $\hat{S}^{2}$ operator.
The spin angular momentum along the $k$ $(k=x,y,z)$ direction can be expressed as~\cite{electrons-spin}
\begin{equation}
    \hat{S_k} = \frac{1}{2}
    \sum_{i=0}^{n-1} \sum_{s,s'} c_{i,s}^\dagger \sigma_{s,s'}^{k} c_{i,s'},
\end{equation}
where $n$ is the number of qubits, $\sigma_{s,s'}^{k}$ denotes the $(s,s')$ element of the Pauli matrix in the $k$ direction, and $c_{i,s}^\dagger$ and $c_{i,s}$  are the fermionic creation and annihilation operators.
The total spin squared operator $\hat{S}^{2}$ can be expressed as the summation of the squared spin angular momentum along the $x,y,z$ directions, written as
\begin{equation}
    \hat{S}^{2} = \hat{S^{2}_{x}} + \hat{S^{2}_{y}} + \hat{S^{2}_{z}}
    = \frac{1}{4} \sum_{k=x,y,z}
    \Bigg(
    \sum_{i=0}^{n-1} \sum_{s,s'} c_{i,s}^\dagger \sigma_{s,s'}^{k} c_{i,s'}
    \Bigg)^2.
\end{equation}
Note that this is the product of two summations, where the index runs over all $n$ qubits. 
Using a naive strategy, the number of Pauli terms in the observable $\hat{S}^{2}$ is $O(n^2)$, which is smaller than the number of Pauli terms $O(n^4)$ in the Hamiltonian $\hat{H}$.

Finally, it is helpful to discuss the present limitations of the VQE algorithm as a means of determining the total spin angular momentum of the ground state.
In order to identify the ground state among the various spin states, the ansatz should suitably represent both the ground state and other electronic states having different spin multiplicities.
In addition, the number of parameters to be optimized should be sufficiently small.
Unfortunately, it is still unclear whether such an ansatz exists.
In this context, developing a sophisticated method for the state preparation of the quantum device will be an important challenge in future.
Assuming that this is possible, the present work indicates that the total spin angular momentum could be measured with a small number of shots.

\newpage

\newpage

\begin{table}[H]
    \centering
    \caption{The potential energies and total spin angular momentum values calculated using the IBM \textit{ibm\_canberra} quantum device, employing three replicate VQE calculations. \textit{r}(Pt--C) denotes the bond length between Pt and C in Å, while $E_{\rm{CASCI}}$, $\ev*{\hat{H}}$, and $e_{\ev*{\hat{H}}}$ denote the reference potential energy calculated using the CASCI method, the expected value of the Hamiltonian, and the associated standard error, respectively, in Hartree. $\ev*{\hat{S}^{2}}_{\rm{CASCI}}$, $\ev*{\hat{S^2}}$, and $e_{\ev*{\hat{S^2}}}$ denote the reference CASCI value for the total spin angular momentum, the expected value for $\hat{S}^{2}$, and the associated standard error, respectively. The three values at each \textit{r}(Pt--C) correspond to the results of three replicate VQE calculations.}
    \begin{tabular}{lcccccc} \hline \hline
       \textit{r}(Pt--C) & $E_{\rm{CASCI}}$ & $\ev*{\hat{H}}$ & $e_{\ev*{\hat{H}}}$ & $\ev*{\hat{S^2}}_{\rm{CASCI}}$ & $\ev*{\hat{S}^{2}}$ & $e_{\ev*{\hat{S^2}}}$ \\ \hline
         1.846 &  -231.1477 & -231.1451  & 0.0009 &  0.0 & -0.06 & 0.19 \\
               &     & -231.1460 & 0.0008 &   & 0.01 & 0.19 \\
               &     & -231.1460 & 0.0008 &   & 0.36 & 0.19 \\
         3.0 &  -231.1364 & -231.1342 & 0.0005 &  2.0 & 1.81 & 0.13 \\ 
            &      & -231.1372 & 0.0005 &   & 2.01 & 0.09 \\ 
            &      & -231.1355 & 0.0005 &   & 1.86 & 0.14 \\ 
         \hline \hline
    \end{tabular}
    \label{tab:ibm_result}
\end{table}

\newpage

\section{Conclusion}
Working within the framework of the VQE quantum-classical hybrid algorithm, the ground state of a molecular system was calculated while minimizing the cost function. 
The energy and spin multiplicity of the ground state could be simultaneously explored when applying this algorithm without penalty terms on the spin angular momentum.  This computational scheme could be useful in the calculation of a strongly correlated system for which the spin multiplicity of the ground state cannot be easily estimated. 
In this study, the adsorption of CO onto a single Pt atom was investigated as a model system. 
VQE calculations using the statevector simulator successfully traced the ground state over the entire potential energy curve. These calculations converged to the singlet state in the bonding region but the triplet state at the dissociation limit. 
In addition, the potential energies acquired using an actual quantum device qualitatively reproduced the CASCI energy values for both the bonding and dissociation regions.
Furthermore, the spin multiplicities in the bonding and dissociation regions could be clearly distinguished in spite of the noise effect. Thus, using the VQE algorithm, it was possible to identify a change in spin multiplicity moving along the ground state potential energy curve without specifying the spin multiplicity in advance.
The results of this research suggest that discrete values of parameters such as spin angular momentum can be determined with reasonable accuracy even using NISQ devices if the ansatz that is adopted has sufficient expressibility to describe the multi-spin states, including the target state.

\subsection{Notes}
The authors declare no competing financial interest.


\bibliography{ref}

\providecommand{\latin}[1]{#1}
\makeatletter
\providecommand{\doi}
  {\begingroup\let\do\@makeother\dospecials
  \catcode`\{=1 \catcode`\}=2 \doi@aux}
\providecommand{\doi@aux}[1]{\endgroup\texttt{#1}}
\makeatother
\providecommand*\mcitethebibliography{\thebibliography}
\csname @ifundefined\endcsname{endmcitethebibliography}
  {\let\endmcitethebibliography\endthebibliography}{}
\begin{mcitethebibliography}{59}
\providecommand*\natexlab[1]{#1}
\providecommand*\mciteSetBstSublistMode[1]{}
\providecommand*\mciteSetBstMaxWidthForm[2]{}
\providecommand*\mciteBstWouldAddEndPuncttrue
  {\def\EndOfBibitem{\unskip.}}
\providecommand*\mciteBstWouldAddEndPunctfalse
  {\let\EndOfBibitem\relax}
\providecommand*\mciteSetBstMidEndSepPunct[3]{}
\providecommand*\mciteSetBstSublistLabelBeginEnd[3]{}
\providecommand*\EndOfBibitem{}
\mciteSetBstSublistMode{f}
\mciteSetBstMaxWidthForm{subitem}{(\alph{mcitesubitemcount})}
\mciteSetBstSublistLabelBeginEnd
  {\mcitemaxwidthsubitemform\space}
  {\relax}
  {\relax}

\bibitem[Höök and Tang(2013)Höök, and Tang]{fossilfuel2013}
Höök,~M.; Tang,~X. Depletion of fossil fuels and anthropogenic climate
  change—A review. \emph{Energy policy} \textbf{2013}, \emph{52},
  797--809\relax
\mciteBstWouldAddEndPuncttrue
\mciteSetBstMidEndSepPunct{\mcitedefaultmidpunct}
{\mcitedefaultendpunct}{\mcitedefaultseppunct}\relax
\EndOfBibitem
\bibitem[Welsby \latin{et~al.}(2021)Welsby, Price, Pye, and Ekins]{climate2021}
Welsby,~D.; Price,~J.; Pye,~S.; Ekins,~P. Unextractable fossil fuels in a 1.5 C
  world. \emph{Nature} \textbf{2021}, \emph{597}, 230--234\relax
\mciteBstWouldAddEndPuncttrue
\mciteSetBstMidEndSepPunct{\mcitedefaultmidpunct}
{\mcitedefaultendpunct}{\mcitedefaultseppunct}\relax
\EndOfBibitem
\bibitem[Yagi and Kaneko(2001)Yagi, and Kaneko]{yagi2001}
Yagi,~M.; Kaneko,~M. Molecular catalysts for water oxidation. \emph{Chem. Rev.}
  \textbf{2001}, \emph{101}, 21--36\relax
\mciteBstWouldAddEndPuncttrue
\mciteSetBstMidEndSepPunct{\mcitedefaultmidpunct}
{\mcitedefaultendpunct}{\mcitedefaultseppunct}\relax
\EndOfBibitem
\bibitem[Kärkäs \latin{et~al.}(2014)Kärkäs, Verho, Johnston, and
  Åkermark]{karkas2014}
Kärkäs,~M.~D.; Verho,~O.; Johnston,~E.~V.; Åkermark,~B. Artificial
  photosynthesis: molecular systems for catalytic water oxidation. \emph{Chem.
  Rev.} \textbf{2014}, \emph{114}, 11863--12001\relax
\mciteBstWouldAddEndPuncttrue
\mciteSetBstMidEndSepPunct{\mcitedefaultmidpunct}
{\mcitedefaultendpunct}{\mcitedefaultseppunct}\relax
\EndOfBibitem
\bibitem[Blakemore \latin{et~al.}(2015)Blakemore, Crabtree, and
  Brudvig]{blakemore2015}
Blakemore,~J.~D.; Crabtree,~R.~H.; Brudvig,~G.~W. Molecular catalysts for water
  oxidation. \emph{Chem. Rev.} \textbf{2015}, \emph{115}, 12974--13005\relax
\mciteBstWouldAddEndPuncttrue
\mciteSetBstMidEndSepPunct{\mcitedefaultmidpunct}
{\mcitedefaultendpunct}{\mcitedefaultseppunct}\relax
\EndOfBibitem
\bibitem[Lloyd(2011)]{lloyd2011handbook}
Lloyd,~L. \emph{Handbook of industrial catalysts}; Springer Science \& Business
  Media, 2011\relax
\mciteBstWouldAddEndPuncttrue
\mciteSetBstMidEndSepPunct{\mcitedefaultmidpunct}
{\mcitedefaultendpunct}{\mcitedefaultseppunct}\relax
\EndOfBibitem
\bibitem[Hagen(2015)]{hagen2015industrial}
Hagen,~J. \emph{Industrial catalysis: a practical approach}; John Wiley \&
  Sons, 2015\relax
\mciteBstWouldAddEndPuncttrue
\mciteSetBstMidEndSepPunct{\mcitedefaultmidpunct}
{\mcitedefaultendpunct}{\mcitedefaultseppunct}\relax
\EndOfBibitem
\bibitem[Sherrill and Schaefer~III(1999)Sherrill, and Schaefer~III]{ci1999}
Sherrill,~C.~D.; Schaefer~III,~H.~F. \emph{Adv. Quantum Chem.}; Elsevier, 1999;
  Vol.~34; pp 143--269\relax
\mciteBstWouldAddEndPuncttrue
\mciteSetBstMidEndSepPunct{\mcitedefaultmidpunct}
{\mcitedefaultendpunct}{\mcitedefaultseppunct}\relax
\EndOfBibitem
\bibitem[Olsen \latin{et~al.}(1996)Olsen, Jo/rgensen, Koch, Balkova, and
  Bartlett]{fci1996}
Olsen,~J.; Jo/rgensen,~P.; Koch,~H.; Balkova,~A.; Bartlett,~R.~J. Full
  configuration--interaction and state of the art correlation calculations on
  water in a valence double-zeta basis with polarization functions. \emph{J.
  Chem. Phys.} \textbf{1996}, \emph{104}, 8007--8015\relax
\mciteBstWouldAddEndPuncttrue
\mciteSetBstMidEndSepPunct{\mcitedefaultmidpunct}
{\mcitedefaultendpunct}{\mcitedefaultseppunct}\relax
\EndOfBibitem
\bibitem[Cao \latin{et~al.}(2019)Cao, Romero, Olson, Degroote, Johnson,
  Kieferov{\'a}, Kivlichan, Menke, Peropadre, Sawaya, \latin{et~al.}
  others]{qcqc2019}
Cao,~Y.; Romero,~J.; Olson,~J.~P.; Degroote,~M.; Johnson,~P.~D.;
  Kieferov{\'a},~M.; Kivlichan,~I.~D.; Menke,~T.; Peropadre,~B.; Sawaya,~N.~P.,
  \latin{et~al.}  Quantum chemistry in the age of quantum computing.
  \emph{Chem. Rev.} \textbf{2019}, \emph{119}, 10856--10915\relax
\mciteBstWouldAddEndPuncttrue
\mciteSetBstMidEndSepPunct{\mcitedefaultmidpunct}
{\mcitedefaultendpunct}{\mcitedefaultseppunct}\relax
\EndOfBibitem
\bibitem[McArdle \latin{et~al.}(2020)McArdle, Endo, Aspuru-Guzik, Benjamin, and
  Yuan]{qcqc2020}
McArdle,~S.; Endo,~S.; Aspuru-Guzik,~A.; Benjamin,~S.~C.; Yuan,~X. Quantum
  computational chemistry. \emph{Rev. Mod. Phys.} \textbf{2020}, \emph{92},
  015003\relax
\mciteBstWouldAddEndPuncttrue
\mciteSetBstMidEndSepPunct{\mcitedefaultmidpunct}
{\mcitedefaultendpunct}{\mcitedefaultseppunct}\relax
\EndOfBibitem
\bibitem[Abrams and Lloyd(1999)Abrams, and Lloyd]{qpe1999}
Abrams,~D.~S.; Lloyd,~S. Quantum algorithm providing exponential speed increase
  for finding eigenvalues and eigenvectors. \emph{Phys. Rev. Lett.}
  \textbf{1999}, \emph{83}, 5162--5165\relax
\mciteBstWouldAddEndPuncttrue
\mciteSetBstMidEndSepPunct{\mcitedefaultmidpunct}
{\mcitedefaultendpunct}{\mcitedefaultseppunct}\relax
\EndOfBibitem
\bibitem[Aspuru-Guzik \latin{et~al.}(2005)Aspuru-Guzik, Dutoi, Love, and
  Head-Gordon]{qpe2005}
Aspuru-Guzik,~A.; Dutoi,~A.~D.; Love,~P.~J.; Head-Gordon,~M. Simulated quantum
  computation of molecular energies. \emph{Science} \textbf{2005}, \emph{309},
  1704--1707\relax
\mciteBstWouldAddEndPuncttrue
\mciteSetBstMidEndSepPunct{\mcitedefaultmidpunct}
{\mcitedefaultendpunct}{\mcitedefaultseppunct}\relax
\EndOfBibitem
\bibitem[Wecker \latin{et~al.}(2014)Wecker, Bauer, Clark, Hastings, and
  Troyer]{qpe2014}
Wecker,~D.; Bauer,~B.; Clark,~B.~K.; Hastings,~M.~B.; Troyer,~M. Gate-count
  estimates for performing quantum chemistry on small quantum computers.
  \emph{Phys. Rev. A} \textbf{2014}, \emph{90}, 022305\relax
\mciteBstWouldAddEndPuncttrue
\mciteSetBstMidEndSepPunct{\mcitedefaultmidpunct}
{\mcitedefaultendpunct}{\mcitedefaultseppunct}\relax
\EndOfBibitem
\bibitem[Babbush \latin{et~al.}(2017)Babbush, Berry, Sanders, Kivlichan,
  Scherer, Wei, Love, and Aspuru-Guzik]{qpe2017}
Babbush,~R.; Berry,~D.~W.; Sanders,~Y.~R.; Kivlichan,~I.~D.; Scherer,~A.;
  Wei,~A.~Y.; Love,~P.~J.; Aspuru-Guzik,~A. Exponentially more precise quantum
  simulation of fermions in the configuration interaction representation.
  \emph{Quantum Sci. Technol.} \textbf{2017}, \emph{3}, 015006\relax
\mciteBstWouldAddEndPuncttrue
\mciteSetBstMidEndSepPunct{\mcitedefaultmidpunct}
{\mcitedefaultendpunct}{\mcitedefaultseppunct}\relax
\EndOfBibitem
\bibitem[Kitagawa \latin{et~al.}(2016)Kitagawa, Chen, Nakatani, Nakayama, and
  Hasegawa]{kitagawa2016dft}
Kitagawa,~Y.; Chen,~Y.; Nakatani,~N.; Nakayama,~A.; Hasegawa,~J.-y. A DFT and
  multi-configurational perturbation theory study on O{$_{2}$} binding to a
  model heme compound via the spin-change barrier. \emph{Phys. Chem. Chem.
  Phys.} \textbf{2016}, \emph{18}, 18137--18144\relax
\mciteBstWouldAddEndPuncttrue
\mciteSetBstMidEndSepPunct{\mcitedefaultmidpunct}
{\mcitedefaultendpunct}{\mcitedefaultseppunct}\relax
\EndOfBibitem
\bibitem[Watanabe \latin{et~al.}(2016)Watanabe, Nakatani, Nakayama, Higashi,
  and Hasegawa]{watanabe2016spin}
Watanabe,~K.; Nakatani,~N.; Nakayama,~A.; Higashi,~M.; Hasegawa,~J.-y.
  Spin-blocking effect in CO and H{$_{2}$} binding reactions to molybdenocene
  and tungstenocene: A theoretical study on the reaction mechanism via the
  minimum energy intersystem crossing point. \emph{Inorg. Chem.} \textbf{2016},
  \emph{55}, 8082--8090\relax
\mciteBstWouldAddEndPuncttrue
\mciteSetBstMidEndSepPunct{\mcitedefaultmidpunct}
{\mcitedefaultendpunct}{\mcitedefaultseppunct}\relax
\EndOfBibitem
\bibitem[Jiang \latin{et~al.}(2016)Jiang, Jiang, and Fu]{jiang2016mechanism}
Jiang,~Y.-Y.; Jiang,~J.-L.; Fu,~Y. Mechanism of vanadium-catalyzed
  deoxydehydration of vicinal diols: Spin-crossover-involved processes.
  \emph{Organometallics} \textbf{2016}, \emph{35}, 3388--3396\relax
\mciteBstWouldAddEndPuncttrue
\mciteSetBstMidEndSepPunct{\mcitedefaultmidpunct}
{\mcitedefaultendpunct}{\mcitedefaultseppunct}\relax
\EndOfBibitem
\bibitem[Kwon \latin{et~al.}(2017)Kwon, Proctor, Mendoza, Uyeda, and
  Ess]{kwon2017catalytic}
Kwon,~D.-H.; Proctor,~M.; Mendoza,~S.; Uyeda,~C.; Ess,~D.~H. Catalytic
  dinuclear nickel spin crossover mechanism and selectivity for alkyne
  cyclotrimerization. \emph{ACS Catal.} \textbf{2017}, \emph{7},
  4796--4804\relax
\mciteBstWouldAddEndPuncttrue
\mciteSetBstMidEndSepPunct{\mcitedefaultmidpunct}
{\mcitedefaultendpunct}{\mcitedefaultseppunct}\relax
\EndOfBibitem
\bibitem[Nakatani \latin{et~al.}(2018)Nakatani, Nakayama, and
  Hasegawa]{Nakatani2018}
Nakatani,~N.; Nakayama,~A.; Hasegawa,~J.-y. In \emph{Frontiers of Quantum
  Chemistry}; W{\'o}jcik,~M.~J., Nakatsuji,~H., Kirtman,~B., Ozaki,~Y., Eds.;
  Springer Singapore: Singapore, 2018; pp 289--313\relax
\mciteBstWouldAddEndPuncttrue
\mciteSetBstMidEndSepPunct{\mcitedefaultmidpunct}
{\mcitedefaultendpunct}{\mcitedefaultseppunct}\relax
\EndOfBibitem
\bibitem[Li \latin{et~al.}(2020)Li, Cao, Hung, Lu, Cai, Rykov, Miao, Xi, Yang,
  Hu, Wang, Zhao, Alp, Xu, Chan, Chen, Xiong, Xiao, Huang, Li, Zhang, and
  Liu]{li2020identification}
Li,~X. \latin{et~al.}  Identification of the electronic and structural dynamics
  of catalytic centers in single-Fe-atom material. \emph{Chem} \textbf{2020},
  \emph{6}, 3440--3454\relax
\mciteBstWouldAddEndPuncttrue
\mciteSetBstMidEndSepPunct{\mcitedefaultmidpunct}
{\mcitedefaultendpunct}{\mcitedefaultseppunct}\relax
\EndOfBibitem
\bibitem[Cramer(2004)]{Christopher2004}
Cramer,~C.~J. \emph{Essentials of Computational Chemistry: Theories and Models,
  2nd Edition}; John Wiley \& Sons, 2004\relax
\mciteBstWouldAddEndPuncttrue
\mciteSetBstMidEndSepPunct{\mcitedefaultmidpunct}
{\mcitedefaultendpunct}{\mcitedefaultseppunct}\relax
\EndOfBibitem
\bibitem[Radon and Pierloot(2008)Radon, and Pierloot]{radon2008binding}
Radon,~M.; Pierloot,~K. Binding of CO, NO, and O{$_{2}$} to heme by density
  functional and multireference ab initio calculations. \emph{J. Phys. Chem. A}
  \textbf{2008}, \emph{112}, 11824--11832\relax
\mciteBstWouldAddEndPuncttrue
\mciteSetBstMidEndSepPunct{\mcitedefaultmidpunct}
{\mcitedefaultendpunct}{\mcitedefaultseppunct}\relax
\EndOfBibitem
\bibitem[Sebetci(2009)]{sebetci2009does}
Sebetci,~A. Does spin--orbit coupling effect favor planar structures for small
  platinum clusters? \emph{Phys. Chem. Chem. Phys.} \textbf{2009}, \emph{11},
  921--925\relax
\mciteBstWouldAddEndPuncttrue
\mciteSetBstMidEndSepPunct{\mcitedefaultmidpunct}
{\mcitedefaultendpunct}{\mcitedefaultseppunct}\relax
\EndOfBibitem
\bibitem[Ali \latin{et~al.}(2012)Ali, Sanyal, and Oppeneer]{ali2012electronic}
Ali,~M.~E.; Sanyal,~B.; Oppeneer,~P.~M. Electronic structure, spin-states, and
  spin-crossover reaction of heme-related Fe-porphyrins: a theoretical
  perspective. \emph{J. Phys. Chem. B} \textbf{2012}, \emph{116},
  5849--5859\relax
\mciteBstWouldAddEndPuncttrue
\mciteSetBstMidEndSepPunct{\mcitedefaultmidpunct}
{\mcitedefaultendpunct}{\mcitedefaultseppunct}\relax
\EndOfBibitem
\bibitem[Garc{\'\i}a-Cruz \latin{et~al.}(2017)Garc{\'\i}a-Cruz, Poulain,
  Hern{\'a}ndez-P{\'e}rez, Reyes-Nava, Gonz{\'a}lez-Torres, Rubio-Ponce, and
  Olvera-Neria]{garcia2017effect}
Garc{\'\i}a-Cruz,~R.; Poulain,~E.; Hern{\'a}ndez-P{\'e}rez,~I.;
  Reyes-Nava,~J.~A.; Gonz{\'a}lez-Torres,~J.~C.; Rubio-Ponce,~A.;
  Olvera-Neria,~O. Effect of Spin Multiplicity in O{$_{2}$} Adsorption and
  Dissociation on Small Bimetallic AuAg Clusters. \emph{J. Phys. Chem. A}
  \textbf{2017}, \emph{121}, 6079--6089\relax
\mciteBstWouldAddEndPuncttrue
\mciteSetBstMidEndSepPunct{\mcitedefaultmidpunct}
{\mcitedefaultendpunct}{\mcitedefaultseppunct}\relax
\EndOfBibitem
\bibitem[Zhang \latin{et~al.}(2020)Zhang, Zhang, Xin, Chen, Hu, Zhang, and
  Zhao]{zhang2020probing}
Zhang,~F.; Zhang,~H.; Xin,~W.; Chen,~P.; Hu,~Y.; Zhang,~X.; Zhao,~Y. Probing
  the structural evolution and electronic properties of divalent metal
  Be{$_{2}$}Mg n clusters from small to medium-size. \emph{Sci. Rep.}
  \textbf{2020}, \emph{10}, 6052\relax
\mciteBstWouldAddEndPuncttrue
\mciteSetBstMidEndSepPunct{\mcitedefaultmidpunct}
{\mcitedefaultendpunct}{\mcitedefaultseppunct}\relax
\EndOfBibitem
\bibitem[Preskill(2018)]{nisq2018}
Preskill,~J. Quantum computing in the NISQ era and beyond. \emph{Quantum}
  \textbf{2018}, \emph{2}, 79\relax
\mciteBstWouldAddEndPuncttrue
\mciteSetBstMidEndSepPunct{\mcitedefaultmidpunct}
{\mcitedefaultendpunct}{\mcitedefaultseppunct}\relax
\EndOfBibitem
\bibitem[Wang \latin{et~al.}(2019)Wang, Higgott, and Brierley]{nisq2019}
Wang,~D.; Higgott,~O.; Brierley,~S. Accelerated variational quantum
  eigensolver. \emph{Phys. Rev. Lett.} \textbf{2019}, \emph{122}, 140504\relax
\mciteBstWouldAddEndPuncttrue
\mciteSetBstMidEndSepPunct{\mcitedefaultmidpunct}
{\mcitedefaultendpunct}{\mcitedefaultseppunct}\relax
\EndOfBibitem
\bibitem[Sugisaki \latin{et~al.}(2021)Sugisaki, Sakai, Toyota, Sato, Shiomi,
  and Takui]{nisq2021}
Sugisaki,~K.; Sakai,~C.; Toyota,~K.; Sato,~K.; Shiomi,~D.; Takui,~T. Bayesian
  phase difference estimation: a general quantum algorithm for the direct
  calculation of energy gaps. \emph{Phys. Chem. Chem. Phys.} \textbf{2021},
  \emph{23}, 20152--20162\relax
\mciteBstWouldAddEndPuncttrue
\mciteSetBstMidEndSepPunct{\mcitedefaultmidpunct}
{\mcitedefaultendpunct}{\mcitedefaultseppunct}\relax
\EndOfBibitem
\bibitem[Peruzzo \latin{et~al.}(2014)Peruzzo, McClean, Shadbolt, Yung, Zhou,
  Love, Aspuru-Guzik, and O’brien]{vqe2014}
Peruzzo,~A.; McClean,~J.; Shadbolt,~P.; Yung,~M.-H.; Zhou,~X.-Q.; Love,~P.~J.;
  Aspuru-Guzik,~A.; O’brien,~J.~L. A variational eigenvalue solver on a
  photonic quantum processor. \emph{Nat. Commun.} \textbf{2014}, \emph{5},
  1--7\relax
\mciteBstWouldAddEndPuncttrue
\mciteSetBstMidEndSepPunct{\mcitedefaultmidpunct}
{\mcitedefaultendpunct}{\mcitedefaultseppunct}\relax
\EndOfBibitem
\bibitem[McClean \latin{et~al.}(2016)McClean, Romero, Babbush, and
  Aspuru-Guzik]{vqe2016}
McClean,~J.~R.; Romero,~J.; Babbush,~R.; Aspuru-Guzik,~A. The theory of
  variational hybrid quantum-classical algorithms. \emph{New J. Phys.}
  \textbf{2016}, \emph{18}, 023023\relax
\mciteBstWouldAddEndPuncttrue
\mciteSetBstMidEndSepPunct{\mcitedefaultmidpunct}
{\mcitedefaultendpunct}{\mcitedefaultseppunct}\relax
\EndOfBibitem
\bibitem[Roszak and Balasubramanian(1993)Roszak, and Balasubramanian]{ptco1993}
Roszak,~S.; Balasubramanian,~K. Potential energy curves for platinum-carbon
  monoxide interactions. \emph{J. Phys. Chem.} \textbf{1993}, \emph{97},
  11238--11241\relax
\mciteBstWouldAddEndPuncttrue
\mciteSetBstMidEndSepPunct{\mcitedefaultmidpunct}
{\mcitedefaultendpunct}{\mcitedefaultseppunct}\relax
\EndOfBibitem
\bibitem[Wu \latin{et~al.}(2004)Wu, Li, Zhang, and Meng]{ptco2004}
Wu,~Z.; Li,~H.; Zhang,~H.; Meng,~J. Electronic structures of MCO (M= Nb, Ta,
  Rh, Ir, Pd, Pt) molecules by density functional theory. \emph{J. Phys. Chem.
  A} \textbf{2004}, \emph{108}, 10906--10910\relax
\mciteBstWouldAddEndPuncttrue
\mciteSetBstMidEndSepPunct{\mcitedefaultmidpunct}
{\mcitedefaultendpunct}{\mcitedefaultseppunct}\relax
\EndOfBibitem
\bibitem[Blaise \latin{et~al.}(1992)Blaise, Verges, Wyart, and
  Engleman]{platinum1992}
Blaise,~J.; Verges,~J.; Wyart,~J.; Engleman,~R.~J. Energy-levels of neutral
  platinum. \emph{J. Res. Natl. Inst} \textbf{1992}, \emph{97}, 213--216\relax
\mciteBstWouldAddEndPuncttrue
\mciteSetBstMidEndSepPunct{\mcitedefaultmidpunct}
{\mcitedefaultendpunct}{\mcitedefaultseppunct}\relax
\EndOfBibitem
\bibitem[Ryabinkin \latin{et~al.}(2019)Ryabinkin, Genin, and
  Izmaylov]{rvqe2019}
Ryabinkin,~I.~G.; Genin,~S.~N.; Izmaylov,~A.~F. Constrained variational quantum
  eigensolver: Quantum computer search engine in the Fock space. \emph{J. Chem.
  Theory Comput.} \textbf{2019}, \emph{15}, 249--255\relax
\mciteBstWouldAddEndPuncttrue
\mciteSetBstMidEndSepPunct{\mcitedefaultmidpunct}
{\mcitedefaultendpunct}{\mcitedefaultseppunct}\relax
\EndOfBibitem
\bibitem[Kuroiwa and Nakagawa(2021)Kuroiwa, and Nakagawa]{kuroiwa2021}
Kuroiwa,~K.; Nakagawa,~Y.~O. Penalty methods for a variational quantum
  eigensolver. \emph{Phys. Rev. Res.} \textbf{2021}, \emph{3}, 013197\relax
\mciteBstWouldAddEndPuncttrue
\mciteSetBstMidEndSepPunct{\mcitedefaultmidpunct}
{\mcitedefaultendpunct}{\mcitedefaultseppunct}\relax
\EndOfBibitem
\bibitem[Vosko \latin{et~al.}(1980)Vosko, Wilk, and Nusair]{b3lyp1980}
Vosko,~S.~H.; Wilk,~L.; Nusair,~M. Accurate spin-dependent electron liquid
  correlation energies for local spin density calculations: a critical
  analysis. \emph{Can. J. Phys.} \textbf{1980}, \emph{58}, 1200--1211\relax
\mciteBstWouldAddEndPuncttrue
\mciteSetBstMidEndSepPunct{\mcitedefaultmidpunct}
{\mcitedefaultendpunct}{\mcitedefaultseppunct}\relax
\EndOfBibitem
\bibitem[Lee \latin{et~al.}(1988)Lee, Yang, and Parr]{b3lyp1988}
Lee,~C.; Yang,~W.; Parr,~R.~G. Development of the Colle-Salvetti
  correlation-energy formula into a functional of the electron density.
  \emph{Phys. Rev. B} \textbf{1988}, \emph{37}, 785--789\relax
\mciteBstWouldAddEndPuncttrue
\mciteSetBstMidEndSepPunct{\mcitedefaultmidpunct}
{\mcitedefaultendpunct}{\mcitedefaultseppunct}\relax
\EndOfBibitem
\bibitem[Becke(1993)]{b3lyp1993}
Becke,~A.~D. Density‐functional thermochemistry. III. The role of exact
  exchange. \emph{J. Chem. Phys.} \textbf{1993}, \emph{98}, 5648--5652\relax
\mciteBstWouldAddEndPuncttrue
\mciteSetBstMidEndSepPunct{\mcitedefaultmidpunct}
{\mcitedefaultendpunct}{\mcitedefaultseppunct}\relax
\EndOfBibitem
\bibitem[Stephens \latin{et~al.}(1994)Stephens, Devlin, Chabalowski, and
  Frisch]{b3lyp1994}
Stephens,~P.~J.; Devlin,~F.~J.; Chabalowski,~C.~F.; Frisch,~M.~J. Ab initio
  calculation of vibrational absorption and circular dichroism spectra using
  density functional force fields. \emph{J. Phys. Chem.} \textbf{1994},
  \emph{98}, 11623--11627\relax
\mciteBstWouldAddEndPuncttrue
\mciteSetBstMidEndSepPunct{\mcitedefaultmidpunct}
{\mcitedefaultendpunct}{\mcitedefaultseppunct}\relax
\EndOfBibitem
\bibitem[Weigend and Ahlrichs(2005)Weigend, and Ahlrichs]{ahlrichs2005}
Weigend,~F.; Ahlrichs,~R. Balanced basis sets of split valence, triple zeta
  valence and quadruple zeta valence quality for H to Rn: Design and assessment
  of accuracy. \emph{Phys. Chem. Chem. Phys.} \textbf{2005}, \emph{7},
  3297--3305\relax
\mciteBstWouldAddEndPuncttrue
\mciteSetBstMidEndSepPunct{\mcitedefaultmidpunct}
{\mcitedefaultendpunct}{\mcitedefaultseppunct}\relax
\EndOfBibitem
\bibitem[Dunning~Jr(1989)]{dunning1989}
Dunning~Jr,~T.~H. Gaussian basis sets for use in correlated molecular
  calculations. I. The atoms boron through neon and hydrogen. \emph{J. Chem.
  Phys.} \textbf{1989}, \emph{90}, 1007--1023\relax
\mciteBstWouldAddEndPuncttrue
\mciteSetBstMidEndSepPunct{\mcitedefaultmidpunct}
{\mcitedefaultendpunct}{\mcitedefaultseppunct}\relax
\EndOfBibitem
\bibitem[Frisch \latin{et~al.}(2016)Frisch, Trucks, Schlegel, Scuseria, Robb,
  Cheeseman, Scalmani, Barone, Petersson, Nakatsuji, \latin{et~al.}
  others]{g16}
Frisch,~M.~J.; Trucks,~G.~W.; Schlegel,~H.~B.; Scuseria,~G.~E.; Robb,~M.~A.;
  Cheeseman,~J.~R.; Scalmani,~G.; Barone,~V.; Petersson,~G.~A.; Nakatsuji,~H.,
  \latin{et~al.}  Gaussian 16 {R}evision {C}.01. 2016; Gaussian Inc.
  Wallingford CT\relax
\mciteBstWouldAddEndPuncttrue
\mciteSetBstMidEndSepPunct{\mcitedefaultmidpunct}
{\mcitedefaultendpunct}{\mcitedefaultseppunct}\relax
\EndOfBibitem
\bibitem[Roos \latin{et~al.}(1980)Roos, Taylor, and Sigbahn]{casscf1980}
Roos,~B.~O.; Taylor,~P.~R.; Sigbahn,~P.~E. A complete active space SCF method
  (CASSCF) using a density matrix formulated super-CI approach. \emph{Chem.
  Phys.} \textbf{1980}, \emph{48}, 157--173\relax
\mciteBstWouldAddEndPuncttrue
\mciteSetBstMidEndSepPunct{\mcitedefaultmidpunct}
{\mcitedefaultendpunct}{\mcitedefaultseppunct}\relax
\EndOfBibitem
\bibitem[Roos(1987)]{casscf1987}
Roos,~B.~O. The complete active space self-consistent field method and its
  applications in electronic structure calculations. \emph{Adv. Chem. Phys.}
  \textbf{1987}, \emph{69}, 399--445\relax
\mciteBstWouldAddEndPuncttrue
\mciteSetBstMidEndSepPunct{\mcitedefaultmidpunct}
{\mcitedefaultendpunct}{\mcitedefaultseppunct}\relax
\EndOfBibitem
\bibitem[Levine \latin{et~al.}(2021)Levine, Durden, Esch, Liang, and
  Shu]{casci2021}
Levine,~B.~G.; Durden,~A.~S.; Esch,~M.~P.; Liang,~F.; Shu,~Y. CAS without
  SCF—Why to use CASCI and where to get the orbitals. \emph{J. Chem. Phys.}
  \textbf{2021}, \emph{154}, 090902\relax
\mciteBstWouldAddEndPuncttrue
\mciteSetBstMidEndSepPunct{\mcitedefaultmidpunct}
{\mcitedefaultendpunct}{\mcitedefaultseppunct}\relax
\EndOfBibitem
\bibitem[Mizukami \latin{et~al.}(2020)Mizukami, Mitarai, Nakagawa, Yamamoto,
  Yan, and Ohnishi]{oovqe2020}
Mizukami,~W.; Mitarai,~K.; Nakagawa,~Y.~O.; Yamamoto,~T.; Yan,~T.;
  Ohnishi,~Y.-y. Orbital optimized unitary coupled cluster theory for quantum
  computer. \emph{Phys. Rev. Res.} \textbf{2020}, \emph{2}, 033421\relax
\mciteBstWouldAddEndPuncttrue
\mciteSetBstMidEndSepPunct{\mcitedefaultmidpunct}
{\mcitedefaultendpunct}{\mcitedefaultseppunct}\relax
\EndOfBibitem
\bibitem[Omiya \latin{et~al.}(2022)Omiya, Nakagawa, Koh, Mizukami, Gao, and
  Kobayashi]{oovqe2022}
Omiya,~K.; Nakagawa,~Y.~O.; Koh,~S.; Mizukami,~W.; Gao,~Q.; Kobayashi,~T.
  Analytical energy gradient for state-averaged orbital-optimized variational
  quantum eigensolvers and its application to a photochemical reaction.
  \emph{J. Chem. Theory Comput.} \textbf{2022}, \emph{18}, 741--748\relax
\mciteBstWouldAddEndPuncttrue
\mciteSetBstMidEndSepPunct{\mcitedefaultmidpunct}
{\mcitedefaultendpunct}{\mcitedefaultseppunct}\relax
\EndOfBibitem
\bibitem[Schmidt \latin{et~al.}(1993)Schmidt, Baldridge, Boatz, Elbert, Gordon,
  Jensen, Koseki, Matsunaga, Nguyen, Su, \latin{et~al.} others]{gamess1993}
Schmidt,~M.~W.; Baldridge,~K.~K.; Boatz,~J.~A.; Elbert,~S.~T.; Gordon,~M.~S.;
  Jensen,~J.~H.; Koseki,~S.; Matsunaga,~N.; Nguyen,~K.~A.; Su,~S.,
  \latin{et~al.}  General atomic and molecular electronic structure system.
  \emph{J. Comput. Chem.} \textbf{1993}, \emph{14}, 1347--1363\relax
\mciteBstWouldAddEndPuncttrue
\mciteSetBstMidEndSepPunct{\mcitedefaultmidpunct}
{\mcitedefaultendpunct}{\mcitedefaultseppunct}\relax
\EndOfBibitem
\bibitem[Gordon and Schmidt(2005)Gordon, and Schmidt]{gamess2005}
Gordon,~M.~S.; Schmidt,~M.~W. \emph{Theory and applications of computational
  chemistry}; Elsevier, 2005; pp 1167--1189\relax
\mciteBstWouldAddEndPuncttrue
\mciteSetBstMidEndSepPunct{\mcitedefaultmidpunct}
{\mcitedefaultendpunct}{\mcitedefaultseppunct}\relax
\EndOfBibitem
\bibitem[Abraham and \emph{et al.}(2021)Abraham, and \emph{et al.}]{Qiskit}
Abraham,~H.; \emph{et al.}, Qiskit: An Open-source Framework for Quantum
  Computing. 2021\relax
\mciteBstWouldAddEndPuncttrue
\mciteSetBstMidEndSepPunct{\mcitedefaultmidpunct}
{\mcitedefaultendpunct}{\mcitedefaultseppunct}\relax
\EndOfBibitem
\bibitem[ibm()]{ibm_quantum}
{IBM Quantum.} \url{ https://quantum-computing.ibm.com/}, 2021\relax
\mciteBstWouldAddEndPuncttrue
\mciteSetBstMidEndSepPunct{\mcitedefaultmidpunct}
{\mcitedefaultendpunct}{\mcitedefaultseppunct}\relax
\EndOfBibitem
\bibitem[Sun \latin{et~al.}(2018)Sun, Berkelbach, Blunt, Booth, Guo, Li, Liu,
  McClain, Sayfutyarova, Sharma, Wouters, and Chan]{sun2018pyscf}
Sun,~Q.; Berkelbach,~T.~C.; Blunt,~N.~S.; Booth,~G.~H.; Guo,~S.; Li,~Z.;
  Liu,~J.; McClain,~J.~D.; Sayfutyarova,~E.~R.; Sharma,~S.; Wouters,~S.;
  Chan,~G. K.-L. PySCF: the Python-based simulations of chemistry framework.
  \emph{Wiley Interdiscip. Rev. Comput. Mol. Sci.} \textbf{2018}, \emph{8},
  e1340\relax
\mciteBstWouldAddEndPuncttrue
\mciteSetBstMidEndSepPunct{\mcitedefaultmidpunct}
{\mcitedefaultendpunct}{\mcitedefaultseppunct}\relax
\EndOfBibitem
\bibitem[Kandala \latin{et~al.}(2017)Kandala, Mezzacapo, Temme, Takita, Brink,
  Chow, and Gambetta]{he2017}
Kandala,~A.; Mezzacapo,~A.; Temme,~K.; Takita,~M.; Brink,~M.; Chow,~J.~M.;
  Gambetta,~J.~M. Hardware-efficient variational quantum eigensolver for small
  molecules and quantum magnets. \emph{Nature} \textbf{2017}, \emph{549},
  242--246\relax
\mciteBstWouldAddEndPuncttrue
\mciteSetBstMidEndSepPunct{\mcitedefaultmidpunct}
{\mcitedefaultendpunct}{\mcitedefaultseppunct}\relax
\EndOfBibitem
\bibitem[Nakanishi \latin{et~al.}(2020)Nakanishi, Fujii, and Todo]{NFT}
Nakanishi,~K.~M.; Fujii,~K.; Todo,~S. Sequential minimal optimization for
  quantum-classical hybrid algorithms. \emph{Phys. Rev. Res.} \textbf{2020},
  \emph{2}, 043158\relax
\mciteBstWouldAddEndPuncttrue
\mciteSetBstMidEndSepPunct{\mcitedefaultmidpunct}
{\mcitedefaultendpunct}{\mcitedefaultseppunct}\relax
\EndOfBibitem
\bibitem[van~den Berg \latin{et~al.}(2022)van~den Berg, Minev, and
  Temme]{van_den_Berg_2022}
van~den Berg,~E.; Minev,~Z.~K.; Temme,~K. Model-free readout-error mitigation
  for quantum expectation values. \emph{Phys. Rev. A} \textbf{2022},
  \emph{105}, 032620\relax
\mciteBstWouldAddEndPuncttrue
\mciteSetBstMidEndSepPunct{\mcitedefaultmidpunct}
{\mcitedefaultendpunct}{\mcitedefaultseppunct}\relax
\EndOfBibitem
\bibitem[Auerbach(1994)]{electrons-spin}
Auerbach,~A. \emph{Interacting Electrons and Quantum Magnetism}; Springer
  Science \& Business Media, 1994\relax
\mciteBstWouldAddEndPuncttrue
\mciteSetBstMidEndSepPunct{\mcitedefaultmidpunct}
{\mcitedefaultendpunct}{\mcitedefaultseppunct}\relax
\EndOfBibitem
\end{mcitethebibliography}

\ifarXiv
    

\section*{Supplementary material}

# Supporting Information:

# Computational analysis of chemical reaction using variational quantum eigensolver algorithm without specifying spin multiplicity

Soichi Shirai,[*,†] Hokuto Iwakiri,[‡] Keita Kanno,[‡] Takahiro Horiba,[†] Keita Omiya,[‡] Hirotoshi Hirai,[†] and Sho Koh[*,‡]

†*Toyota Central Research & Development Laboratories, Inc.,*
*41-1 Yokomichi, Nagakute, Aichi 480-1192, Japan*
‡*QunaSys Inc., Aqua Hakusan Building 9F,*
*1-13-7 Hakusan, Bunkyo, Tokyo 113-0001, Japan*

E-mail: shirai@mosk.tytlabs.co.jp; koh@qunasys.com

Table S1: Electronic state total energies in Hartree calculated using CASSCF with CAS(10e, 6o). The basis sets used were def2-QZVP for Pt and cc-pVQZ for C and O. The results were plotted in Figure 4(a) in the manuscript.

| $r$(Pt–C) | $1^1A_1$ | $2^1A_1$ | $3^1A_1$ | $1^3A_1$ | $2^3A_1$ |
|---|---|---|---|---|---|
| 1.55 | -231.149596 | -231.041981 | -231.038449 | -231.070795 | -231.054113 |
| 1.60 | -231.171696 | -231.072580 | -231.071127 | -231.101627 | -231.087124 |
| 1.65 | -231.186850 | -231.095927 | -231.095278 | -231.125182 | -231.112250 |
| 1.70 | -231.196561 | -231.114448 | -231.111660 | -231.142908 | -231.131087 |
| 1.75 | -231.202061 | -231.128006 | -231.123048 | -231.156001 | -231.144947 |
| 1.80 | -231.204358 | -231.137683 | -231.130534 | -231.165448 | -231.154910 |
| 1.85 | -231.204279 | -231.144364 | -231.135027 | -231.172062 | -231.161859 |
| 1.90 | -231.202494 | -231.148773 | -231.137271 | -231.176508 | -231.166515 |
| 1.95 | -231.199546 | -231.151497 | -231.137879 | -231.179328 | -231.169461 |
| 2.00 | -231.195868 | -231.153003 | -231.137346 | -231.180955 | -231.171164 |
| 2.05 | -231.191804 | -231.153662 | -231.136072 | -231.181734 | -231.171991 |
| 2.10 | -231.187621 | -231.153758 | -231.134374 | -231.181931 | -231.172230 |
| 2.15 | -231.183516 | -231.153508 | -231.132505 | -231.181750 | -231.172098 |
| 2.20 | -231.179638 | -231.153071 | -231.130658 | -231.181342 | -231.171759 |
| 2.30 | -231.172901 | -231.152046 | -231.127585 | -231.180256 | -231.170894 |
| 2.40 | -231.167705 | -231.151200 | -231.125884 | -231.179197 | -231.170209 |
| 2.50 | -231.163838 | -231.150740 | -231.125753 | -231.178377 | -231.169932 |
| 2.60 | -231.160950 | -231.150707 | -231.126893 | -231.177843 | -231.170098 |
| 2.70 | -231.158801 | -231.151039 | -231.128741 | -231.177570 | -231.170641 |
| 2.80 | -231.157279 | -231.151624 | -231.130767 | -231.177506 | -231.171449 |
| 2.90 | -231.156301 | -231.152345 | -231.132634 | -231.177594 | -231.172406 |
| 3.00 | -231.155778 | -231.153113 | -231.134184 | -231.177785 | -231.173412 |
| 3.20 | -231.155680 | -231.154569 | -231.136256 | -231.178317 | -231.175300 |
| 3.40 | -231.156197 | -231.155759 | -231.137308 | -231.178869 | -231.176820 |
| 3.60 | -231.156818 | -231.156643 | -231.137810 | -231.179334 | -231.177935 |
| 3.80 | -231.157337 | -231.157264 | -231.138048 | -231.179681 | -231.178711 |
| 4.00 | -231.157713 | -231.157682 | -231.138161 | -231.179920 | -231.179234 |
| 4.20 | -231.157967 | -231.157955 | -231.138212 | -231.180075 | -231.179577 |
| 4.40 | -231.158131 | -231.158126 | -231.138235 | -231.180168 | -231.179800 |
| 4.60 | -231.158234 | -231.158233 | -231.138241 | -231.180220 | -231.179943 |
| 4.80 | -231.158300 | -231.158295 | -231.138241 | -231.180247 | -231.180035 |
| 5.00 | -231.158339 | -231.158333 | -231.138238 | -231.180258 | -231.180094 |

Table S2: Electronic state total energies in Hartree calculated using CASSCF with CAS(4e, 3o). The basis sets used were def2-QZVP for Pt and cc-pVQZ for C and O. The results for $1^1A_1$ and $1^3A_1$ were plotted in Figure S3.

| $r$(Pt–C) | $1^1A_1$ | $2^1A_1$ | $3^1A_1$ | $1^3A_1$ | $2^3A_1$ |
|---|---|---|---|---|---|
| 1.55 | -231.149208 | -231.042007 | -231.038316 | -231.070795 | -231.054113 |
| 1.60 | -231.171258 | -231.072609 | -231.070939 | -231.101627 | -231.087124 |
| 1.65 | -231.186355 | -231.095672 | -231.095310 | -231.125182 | -231.112250 |
| 1.70 | -231.195999 | -231.114113 | -231.111697 | -231.142908 | -231.131087 |
| 1.75 | -231.201422 | -231.127577 | -231.123090 | -231.156001 | -231.144947 |
| 1.80 | -231.203632 | -231.137145 | -231.130583 | -231.165448 | -231.154910 |
| 1.85 | -231.203454 | -231.143701 | -231.135081 | -231.172062 | -231.161859 |
| 1.90 | -231.201558 | -231.147970 | -231.137332 | -231.176508 | -231.166515 |
| 1.95 | -231.198485 | -231.150535 | -231.137943 | -231.179328 | -231.169461 |
| 2.00 | -231.194668 | -231.151865 | -231.137411 | -231.180955 | -231.171164 |
| 2.05 | -231.190443 | -231.152331 | -231.136133 | -231.181734 | -231.171991 |
| 2.10 | -231.186082 | -231.152216 | -231.134422 | -231.181931 | -231.172230 |
| 2.15 | -231.181778 | -231.151739 | -231.132528 | -231.181750 | -231.172098 |
| 2.20 | -231.177677 | -231.151059 | -231.130637 | -231.181342 | -231.171759 |
| 2.30 | -231.170435 | -231.149510 | -231.127391 | -231.180256 | -231.170894 |
| 2.40 | -231.164695 | -231.148110 | -231.125353 | -231.179197 | -231.170209 |
| 2.50 | -231.160357 | -231.147081 | -231.124695 | -231.178377 | -231.169932 |
| 2.60 | -231.157122 | -231.146489 | -231.125216 | -231.177843 | -231.170098 |
| 2.70 | -231.154714 | -231.146305 | -231.126506 | -231.177570 | -231.170641 |
| 2.80 | -231.152948 | -231.146443 | -231.128139 | -231.177506 | -231.171449 |
| 2.90 | -231.151706 | -231.146801 | -231.129796 | -231.177594 | -231.172406 |
| 3.00 | -231.150899 | -231.147288 | -231.131281 | -231.177785 | -231.173412 |
| 3.20 | -231.150246 | -231.148384 | -231.133473 | -231.178317 | -231.175300 |
| 3.40 | -231.150346 | -231.149392 | -231.134720 | -231.178869 | -231.176820 |
| 3.60 | -231.150714 | -231.150199 | -231.135362 | -231.179334 | -231.177935 |
| 3.80 | -231.151095 | -231.150795 | -231.135683 | -231.179681 | -231.178711 |
| 4.00 | -231.151399 | -231.151213 | -231.135844 | -231.179920 | -231.179234 |
| 4.20 | -231.151615 | -231.151494 | -231.135921 | -231.180075 | -231.179577 |
| 4.40 | -231.151759 | -231.151678 | -231.135958 | -231.180168 | -231.179800 |
| 4.60 | -231.151851 | -231.151796 | -231.135973 | -231.180220 | -231.179943 |
| 4.80 | -231.151908 | -231.151871 | -231.135978 | -231.180247 | -231.180035 |
| 5.00 | -231.151942 | -231.151918 | -231.135978 | -231.180258 | -231.180094 |

Table S3: Electronic state total energies in Hartree calculated using CASSCF with CAS(2e, 2o). The basis sets used were def2-QZVP for Pt and cc-pVQZ for C and O. The results for $1^1A_1$ and $1^3A_1$ were plotted in Figure S3.

| $r$(Pt–C) | $1^1A_1$ | $2^1A_1$ | $1^3A_1$ |
|---|---|---|---|
| 1.55 | -231.157046 | -231.040650 | -231.073862 |
| 1.60 | -231.179058 | -231.071036 | -231.104377 |
| 1.65 | -231.194118 | -231.093555 | -231.127693 |
| 1.70 | -231.203719 | -231.109793 | -231.145244 |
| 1.75 | -231.209082 | -231.121070 | -231.158213 |
| 1.80 | -231.211209 | -231.128481 | -231.167573 |
| 1.85 | -231.210920 | -231.132931 | -231.174127 |
| 1.90 | -231.208878 | -231.135166 | -231.178534 |
| 1.95 | -231.205622 | -231.135791 | -231.181326 |
| 2.00 | -231.201584 | -231.135297 | -231.182935 |
| 2.05 | -231.197107 | -231.134075 | -231.183701 |
| 2.10 | -231.192459 | -231.132431 | -231.183888 |
| 2.15 | -231.187849 | -231.130600 | -231.183700 |
| 2.20 | -231.183428 | -231.128762 | -231.183287 |
| 2.30 | -231.175565 | -231.125527 | -231.182196 |
| 2.40 | -231.169305 | -231.123317 | -231.181144 |
| 2.50 | -231.164555 | -231.122329 | -231.180346 |
| 2.60 | -231.160953 | -231.122475 | -231.179852 |
| 2.70 | -231.158154 | -231.123470 | -231.179633 |
| 2.80 | -231.155943 | -231.124959 | -231.179630 |
| 2.90 | -231.154198 | -231.126638 | -231.179780 |
| 3.00 | -231.152825 | -231.128314 | -231.180027 |
| 3.20 | -231.150911 | -231.131264 | -231.180643 |
| 3.40 | -231.149748 | -231.133495 | -231.181247 |
| 3.60 | -231.149040 | -231.135069 | -231.181743 |
| 3.80 | -231.148595 | -231.136144 | -231.182108 |
| 4.00 | -231.148300 | -231.136868 | -231.182359 |
| 4.20 | -231.148099 | -231.137348 | -231.182520 |
| 4.40 | -231.147954 | -231.137668 | -231.182619 |
| 4.60 | -231.147843 | -231.137883 | -231.182675 |
| 4.80 | -231.147760 | -231.138027 | -231.182704 |
| 5.00 | -231.147696 | -231.138125 | -231.182718 |

Table S4: Electronic state total energies in Hartree calculated using CASCI with CAS(2e, 2o). The basis sets used were def2-SVP for Pt and cc-pVDZ for C and O. The results based on ROHF orbital are plotted in Figure 4(b) in the manuscript, while the results based on RHF orbital are plotted in Figure S4.

| $r$(Pt–C) | ROHF orbital | | | RHF orbital | | |
| --- | --- | --- | --- | --- | --- | --- |
| | $1^1A_1$ | $2^1A_1$ | $1^3A_1$ | $1^1A_1$ | $2^1A_1$ | $1^3A_1$ |
| 1.55 | -231.085637 | -230.974423 | -231.010650 | -231.102427 | -230.952334 | -230.973190 |
| 1.60 | -231.109661 | -231.007314 | -231.043878 | -231.127512 | -230.983853 | -231.003618 |
| 1.65 | -231.126273 | -231.031333 | -231.069256 | -231.144947 | -231.006962 | -231.026228 |
| 1.70 | -231.137157 | -231.048365 | -231.088407 | -231.156405 | -231.023428 | -231.042707 |
| 1.75 | -231.143672 | -231.059977 | -231.102656 | -231.163243 | -231.034716 | -231.054434 |
| 1.80 | -231.146911 | -231.067459 | -231.113087 | -231.166565 | -231.042028 | -231.062528 |
| 1.85 | -231.147745 | -231.071857 | -231.120574 | -231.167263 | -231.046350 | -231.067889 |
| 1.90 | -231.146870 | -231.074012 | -231.125824 | -231.166050 | -231.048480 | -231.071238 |
| 1.95 | -231.144829 | -231.074594 | -231.129402 | -231.163493 | -231.049060 | -231.073142 |
| 2.00 | -231.142045 | -231.074131 | -231.131753 | -231.160042 | -231.048599 | -231.074043 |
| 2.05 | -231.138844 | -231.073031 | -231.133224 | -231.156044 | -231.047495 | -231.074279 |
| 2.10 | -231.135468 | -231.071603 | -231.134082 | -231.151765 | -231.046056 | -231.074102 |
| 2.15 | -231.132095 | -231.070077 | -231.134529 | -231.147407 | -231.044511 | -231.073695 |
| 2.20 | -231.128851 | -231.068614 | -231.134714 | -231.143116 | -231.043025 | -231.073187 |
| 2.30 | -231.123061 | -231.066265 | -231.134701 | -231.135119 | -231.040644 | -231.072177 |
| 2.40 | -231.118436 | -231.064935 | -231.134573 | -231.128239 | -231.039361 | -231.071425 |
| 2.50 | -231.115006 | -231.064574 | -231.134556 | -231.122605 | -231.039187 | -231.071007 |
| 2.60 | -231.112634 | -231.064930 | -231.134713 | -231.118156 | -231.039897 | -231.070882 |
| 2.70 | -231.111112 | -231.065713 | -231.135024 | -231.114738 | -231.041182 | -231.070981 |
| 2.80 | -231.110222 | -231.066679 | -231.135439 | -231.112168 | -231.042755 | -231.071234 |
| 2.90 | -231.109770 | -231.067663 | -231.135904 | -231.110268 | -231.044396 | -231.071588 |
| 3.00 | -231.109601 | -231.068567 | -231.136374 | -231.108880 | -231.045963 | -231.071999 |
| 3.20 | -231.109691 | -231.069989 | -231.137215 | -231.107143 | -231.048606 | -231.072872 |
| 3.40 | -231.109963 | -231.070894 | -231.137846 | -231.106226 | -231.050507 | -231.073666 |
| 3.60 | -231.110218 | -231.071409 | -231.138267 | -231.105727 | -231.051769 | -231.074287 |
| 3.80 | -231.110401 | -231.071675 | -231.138525 | -231.105441 | -231.052554 | -231.074710 |
| 4.00 | -231.110515 | -231.071797 | -231.138666 | -231.105264 | -231.053002 | -231.074949 |
| 4.20 | -231.110572 | -231.071839 | -231.138730 | -231.105143 | -231.053223 | -231.075041 |
| 4.40 | -231.110589 | -231.071838 | -231.138744 | -231.105051 | -231.053301 | -231.075029 |
| 4.60 | -231.110577 | -231.071814 | -231.138727 | -231.104972 | -231.053295 | -231.074954 |
| 4.80 | -231.110548 | -231.071779 | -231.138693 | -231.104902 | -231.053250 | -231.074854 |
| 5.00 | -231.110512 | -231.071739 | -231.138651 | -231.104839 | -231.053195 | -231.074755 |

Table S5: Electronic state total energies in Hartree calculated using CASCI. The active space adopted was CAS(2e, 2o). The basis sets used were def2-TZVP for Pt and cc-pVTZ for C and O as "triple zeta", while def2-QZVP for Pt and cc-pVQZ for C and O as "quadruple zeta". The results are plotted in Figure S5.

| $r$(Pt–C) | triple zeta | | | quadruple zeta | | |
|---|---|---|---|---|---|---|
| | $1^1A_1$ | $2^1A_1$ | $1^3A_1$ | $1^1A_1$ | $2^1A_1$ | $1^3A_1$ |
| 1.55 | -231.129257 | -231.022832 | -231.059936 | -231.142662 | -231.036251 | -231.073862 |
| 1.60 | -231.151268 | -231.053887 | -231.091156 | -231.163835 | -231.066609 | -231.104377 |
| 1.65 | -231.166442 | -231.076567 | -231.115026 | -231.178327 | -231.088720 | -231.127693 |
| 1.70 | -231.176294 | -231.092605 | -231.133013 | -231.187624 | -231.104290 | -231.145244 |
| 1.75 | -231.182053 | -231.103451 | -231.146325 | -231.192931 | -231.114745 | -231.158213 |
| 1.80 | -231.184714 | -231.110303 | -231.155959 | -231.195226 | -231.121261 | -231.167574 |
| 1.85 | -231.185084 | -231.114143 | -231.162734 | -231.195303 | -231.124805 | -231.174128 |
| 1.90 | -231.183811 | -231.115768 | -231.167322 | -231.193797 | -231.126163 | -231.178534 |
| 1.95 | -231.181413 | -231.115822 | -231.170267 | -231.191220 | -231.125969 | -231.181327 |
| 2.00 | -231.178300 | -231.114818 | -231.172007 | -231.187973 | -231.124727 | -231.182936 |
| 2.05 | -231.174791 | -231.113160 | -231.172887 | -231.184368 | -231.122836 | -231.183702 |
| 2.10 | -231.171127 | -231.111161 | -231.173179 | -231.180641 | -231.120604 | -231.183889 |
| 2.15 | -231.167487 | -231.109056 | -231.173087 | -231.176964 | -231.118262 | -231.183701 |
| 2.20 | -231.163999 | -231.107018 | -231.172765 | -231.173460 | -231.115982 | -231.183288 |
| 2.30 | -231.157786 | -231.103563 | -231.171842 | -231.167261 | -231.112027 | -231.182197 |
| 2.40 | -231.152817 | -231.101248 | -231.170939 | -231.162345 | -231.109209 | -231.181144 |
| 2.50 | -231.149108 | -231.100072 | -231.170268 | -231.158702 | -231.107560 | -231.180347 |
| 2.60 | -231.146506 | -231.099802 | -231.169876 | -231.156168 | -231.106881 | -231.179853 |
| 2.70 | -231.144799 | -231.100136 | -231.169735 | -231.154520 | -231.106891 | -231.179634 |
| 2.80 | -231.143764 | -231.100807 | -231.169792 | -231.153531 | -231.107324 | -231.179631 |
| 2.90 | -231.143208 | -231.101617 | -231.169991 | -231.153004 | -231.107972 | -231.179781 |
| 3.00 | -231.142977 | -231.102440 | -231.170282 | -231.152784 | -231.108695 | -231.180028 |
| 3.20 | -231.143056 | -231.103876 | -231.170977 | -231.152835 | -231.110057 | -231.180644 |
| 3.40 | -231.143424 | -231.104910 | -231.171642 | -231.153138 | -231.111109 | -231.181248 |
| 3.60 | -231.143815 | -231.105579 | -231.172171 | -231.153463 | -231.111832 | -231.181743 |
| 3.80 | -231.144132 | -231.105983 | -231.172546 | -231.153733 | -231.112296 | -231.182109 |
| 4.00 | -231.144361 | -231.106214 | -231.172792 | -231.153933 | -231.112580 | -231.182359 |
| 4.20 | -231.144513 | -231.106341 | -231.172944 | -231.154069 | -231.112747 | -231.182521 |
| 4.40 | -231.144609 | -231.106408 | -231.173031 | -231.154156 | -231.112844 | -231.182619 |
| 4.60 | -231.144664 | -231.106442 | -231.173077 | -231.154208 | -231.112899 | -231.182676 |
| 4.80 | -231.144694 | -231.106459 | -231.173098 | -231.154237 | -231.112931 | -231.182705 |
| 5.00 | -231.144707 | -231.106467 | -231.173103 | -231.154251 | -231.112951 | -231.182718 |

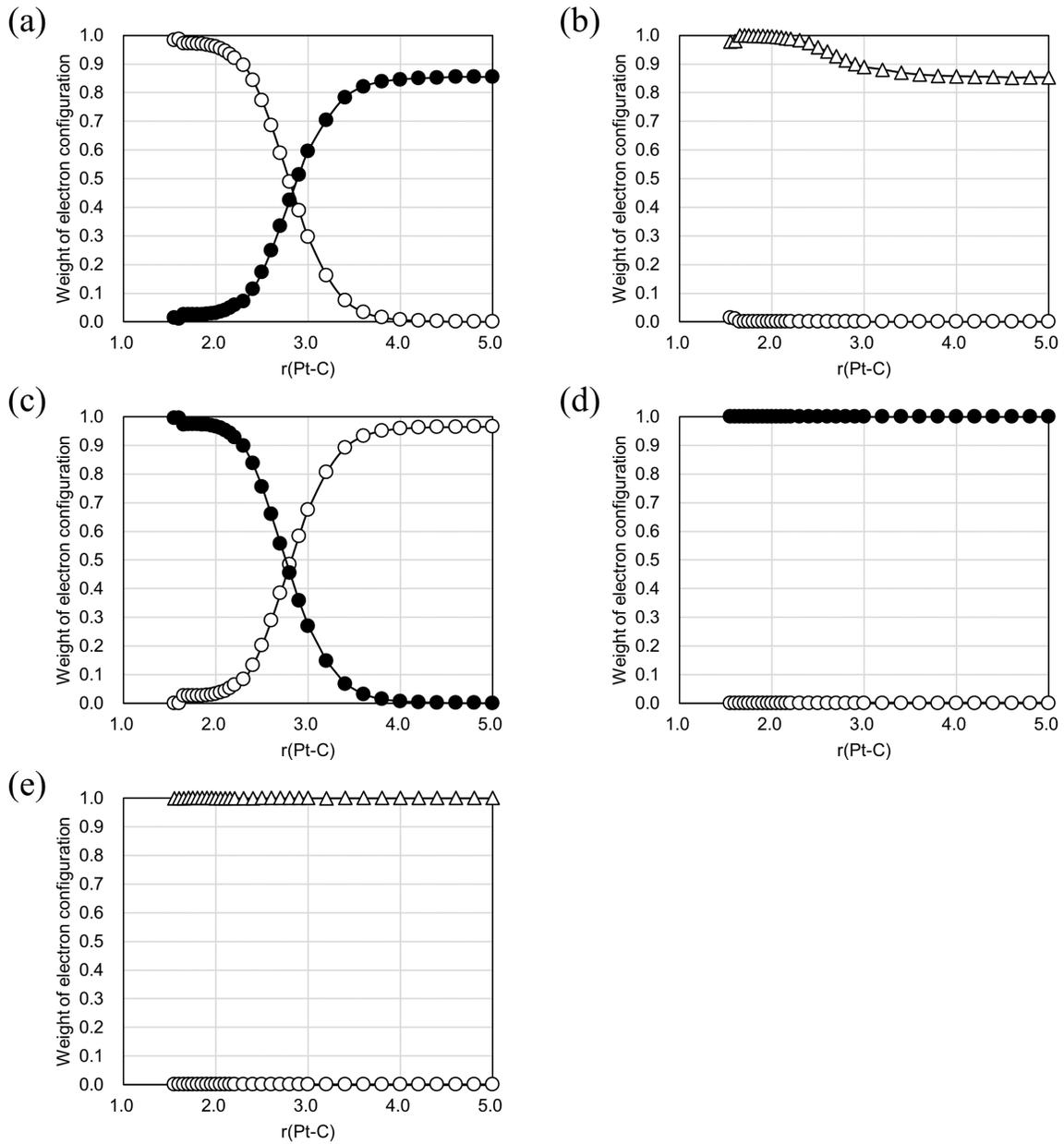

Figure S1: Weights of the electron configurations of $(5d_{xy})^2(5d_{yz})^2(5d_{xz})^2(5d_{x^2-y^2})^2(5d_{z^2})^1(6s)^1$ (filled circle), $(5d_{xy})^2(5d_{yz})^2(5d_{xz})^2(5d_{x^2-y^2})^2(5d_{z^2})^2(6s)^0$ (open circle), and $(5d_{xy})^2(5d_{yz})^2(5d_{xz})^2(5d_{x^2-y^2})^1(5d_{z^2})^2(6s)^1$ (open triangle) for (a) $1^1A_1$, (b) $2^1A_1$, (c) $3^1A_1$, (d) $1^3A_1$, and (e) $2^3A_1$. The weights were calculated as squared coefficients of the configurations in the CASSCF wavefunctions with CAS(10e, 6o).

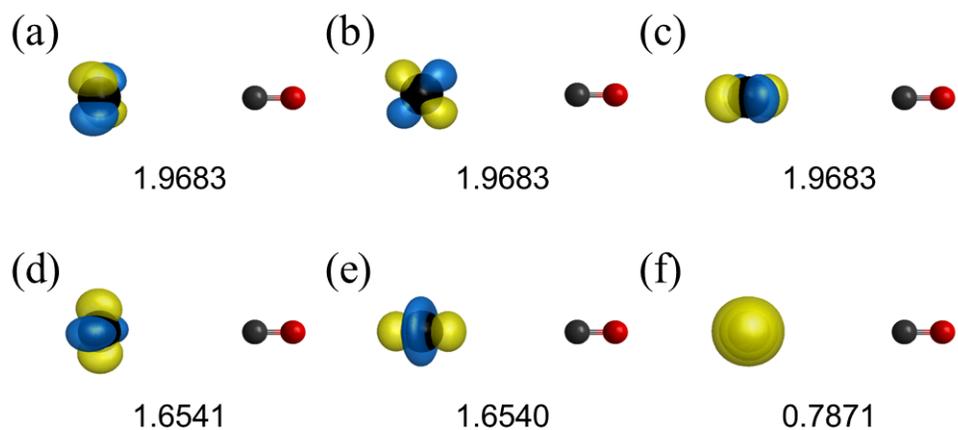

Figure S2: Natural orbitals at $r(\text{Pt–C}) = 5.00$ Å for the singlet states calculated using CASSCF with CAS(10e, 6o). The orbitals are derived from (a) $5d_{xy}$, (b) $5d_{yz}$, (c) $5d_{xz}$, (d) $5d_{x^2-y^2}$, (e) $5d_{z^2}$, and (f) $6s$ of Pt; the symmetries are $a_2$, $b_2$, $b_1$, $a_1$, $a_1$, and $a_1$, respectively. The occupation numbers are presented below the orbitals.

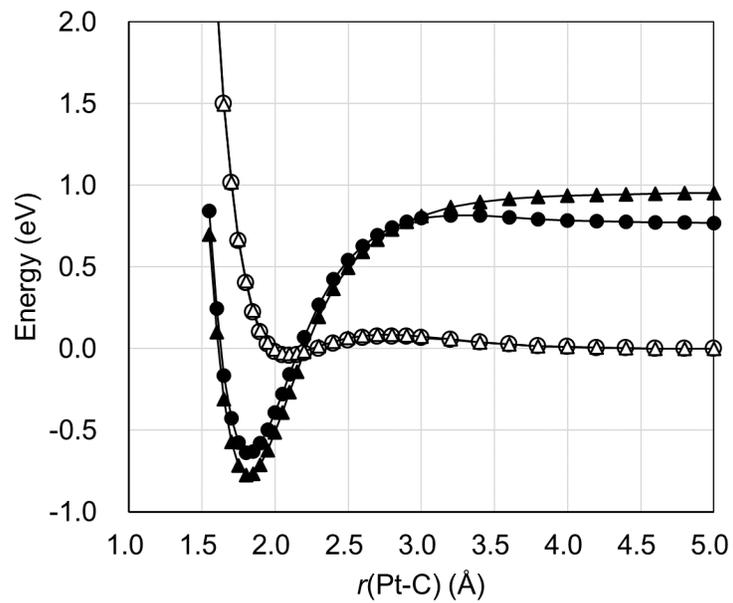

Figure S3: Potential energy curves calculated using CASSCF for $1^1A_1$ with CAS(4e,3o) (filled circle), $1^3A_1$ with CAS(4e,3o) (open circle), $1^1A_1$ with CAS(2e,2o) (filled triangle), and $1^3A_1$ with CAS(2e,2o) (open triangle). The basis set used were def2-QZVP for Pt and cc-pVQZ for C and O.

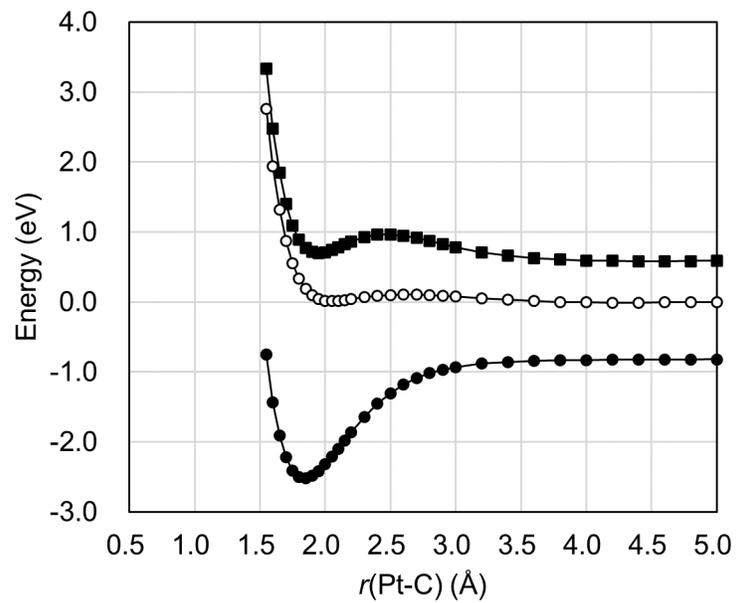

Figure S4: Potential energy curves for $1^1A_1$ (filled circle), $2^1A_1$ (filled square), and $1^3A_1$ (open circle) calculated using CASCI with CAS(2e, 2o) based on the RHF orbitals. The basis set used were def2-SVP for Pt and cc-pVDZ for C and O.


\fi

\end{document}